\documentclass[journal,twoside,web]{ieeecolor}
\usepackage{tmi}
\usepackage{cite}
\usepackage{amsmath,amssymb,amsfonts}
\usepackage{algorithmic}
\usepackage{graphicx}
\usepackage{textcomp}
\usepackage{graphicx}
 \usepackage{amssymb}
 \usepackage{algorithm}
 \usepackage{url}
 \usepackage{subfigure}
  \usepackage{subfig}
  \usepackage{setspace}
  \usepackage{url}
  \usepackage{graphicx}
\usepackage{amssymb}
\usepackage{booktabs}
\usepackage{url}
\usepackage{subfigure}
\usepackage{setspace}
\usepackage{amsmath}
\usepackage{multirow}
\usepackage{algorithm}
\usepackage{algorithmic}
\usepackage[table,xcdraw]{xcolor}
\usepackage{rotating}         
\usepackage{overpic}
\def\BibTeX{{\rm B\kern-.05em{\sc i\kern-.025em b}\kern-.08em
    T\kern-.1667em\lower.7ex\hbox{E}\kern-.125emX}}
\begin{document}
\title{Prototype Learning Guided Hybrid Network for Breast Tumor Segmentation in DCE-MRI}
\author{Lei~Zhou, Yuzhong~Zhang, Jiadong~Zhang, Xuejun Qian, Chen Gong, Kun Sun, Zhongxiang Ding, Xing Wang, Zhenhui Li, Zaiyi~Liu, and~Dinggang~Shen,~\IEEEmembership{Fellow,~IEEE}
\thanks{This work was supported in part by National Natural Science Foundation of China (grant numbers 62131015, 62250710165, U23A20295, 61906121, 82300122), the STI 2030-Major Projects (No. 2022ZD0209000), Shanghai Municipal Central Guided Local Science and Technology Development Fund (grant number YDZX20233100001001), Science and Technology Commission of Shanghai Municipality (STCSM) (grant number 21010502600), and The Key R$\&$D Program of Guangdong Province, China (grant numbers 2023B0303040001, 2021B0101420006).}
\thanks{Lei Zhou and Yuzhong Zhang are with the School of Health Science and Engineering, University of Shanghai for Science and Technology, Shanghai, China (emails: davidzhou@usst.edu.cn,yuzhongzzz1@163.com).}
\thanks{Jiadong Zhang and Xuejun Qian are with the School of Biomedical Engineering, ShanghaiTech University (email: zhangjd1@shanghaitech.edu.cn, qianxj@shanghaitech.edu.cn).}
\thanks{Chen Gong is with the School of Computer Science and Engineering, Nanjing University of Science and Technology (email: chen.gong@njust.edu.cn).}
\thanks{Kun Sun is with Ruijin Hospital, Shanghai Jiaotong University School of Medicine (email: sk12177@rjh.com.cn).}
\thanks{Zhongxiang Ding is with Department of Radiology, Key Laboratory of Clinical Cancer Pharmacology and Toxicology Research of Zhejiang Province (email: hangzhoudzx73@126.com).}
\thanks{Xing Wang is with the Shanghai General Hospital, Shanghai JiaoTong University (email: drwangxing@sjtu.edu.cn.)}
\thanks{Zhenhui Li is with Department of Radiology, The Third Affiliated Hospital of Kunming Medical University (email: lizhenhui@kmmu.edu.cn).}
\thanks{Zaiyi Liu is with Department of Radiology, Guangdong Provincial People's Hospital, Guangdong Academy of Medical Sciences, Guangzhou, 510080, China, and also with Guangdong Provincial Key Laboratory of Artificial Intelligence in Medical Image Analysis and Application, Guangdong Provincial People's Hospital, Guangdong Academy of Medical Sciences, Guangzhou 510080, China (email: liuzaiyi@gdph.org.cn).}
\thanks{D. Shen is with School of Biomedical Engineering $\&$ State Key Laboratory of Advanced Medical Materials and Devices, ShanghaiTech University, Shanghai 201210, China. He is also with Shanghai United Imaging Intelligence Co., Ltd., Shanghai 200232, China, and Shanghai Clinical Research and Trial Center, Shanghai, 201210, China. (e-mail: Dinggang.Shen@gmail.com).}
\thanks{Corresponding author: Lei Zhou and Dinggang Shen.}
}
\maketitle
\begin{abstract}
Automated breast tumor segmentation on the basis of dynamic contrast-enhancement magnetic resonance imaging (DCE-MRI) has shown great promise in clinical practice, particularly for identifying the presence of breast disease. However, accurate segmentation of breast tumor is a challenging task, often necessitating the development of complex networks. To strike an optimal trade-off between computational costs and segmentation performance, we propose a hybrid network via the combination of convolution neural network (CNN) and transformer layers. Specifically, the hybrid network consists of a encoder-decoder architecture by stacking convolution and deconvolution layers. Effective 3D transformer layers are then implemented after the encoder subnetworks, to capture global dependencies between the bottleneck features. To improve the efficiency of hybrid network, two parallel encoder subnetworks are designed for the decoder and the transformer layers, respectively. To further enhance the discriminative capability of hybrid network, a prototype learning guided prediction module is proposed, where the category-specified prototypical features are calculated through online clustering. All learned prototypical features are finally combined with the features from decoder for tumor mask prediction. The experimental results on private and public DCE-MRI datasets demonstrate that the proposed hybrid network achieves superior performance than the state-of-the-art (SOTA) methods, while maintaining balance between segmentation accuracy and computation cost. Moreover, we demonstrate that automatically generated tumor masks can be effectively applied to identify HER2-positive subtype from HER2-negative subtype with the similar accuracy to the analysis based on manual tumor segmentation. The source code is available at https://github.com/ZhouL-lab/PLHN.
\end{abstract}
\vspace{-0.5cm}
\section{Introduction}
\indent \IEEEPARstart{B}{reast} cancer has become the leading cause of cancer death among the women group. The existing literatures reveal that early identification of malignant tumors with timely clinical intervention will considerably improve patient survival and outcomes  \cite{ginsburg2020breast}. Compared with conventional mammography \cite{gotzsche2013screening} and ultrasound imaging \cite{qian2021prospective}, magnetic resonance image (MRI) \cite{van2006magnetic} is a highly sensitive modality for detecting breast cancer with reported sensitivity of more than 80$\%$. Dynamic contrast-enhanced magnetic resonance image (DCE-MRI) has been found to provide excellent performance for tumor detection, lesion diagnosis and even the evaluation of subsequent surgical/treatment plans due to excellent contrast resolution \cite{zhang2023recent}. Artificial intelligence has demonstrated an enormous impact on medical image analysis, including registration \cite{fan2018adversarial}, segmentation \cite{dolz2020deep}, diagnosis \cite{jie2018integration} and reconstruction \cite{bahrami2016convolutional} in recent years. Those techniques can also been applied for breast cancer diagnosis. As a prerequisite of excellent prediction, extracting accurate regions of breast tumors is important while time-consuming because the tumor voxels only occupy a small part of region. Hence, investigating a robust algorithm for breast tumor segmentation is a crucial procedure for promoting the development of breast cancer diagnosis systems.\par
\indent In recent years, several deep learning-based approaches \cite{zhang2018hierarchical,li2019learning,lv2021temporal,wang2021breast,wang2021mixed,zhou2022three} have been proposed to tackle the challenging task of breast tumor segmentation in DCE-MRI images. The first category of methods focuses on designing innovative network architectures. For instance, in \cite{li2019learning}, a multi-stream fusion network was devised to selectively combine valuable information from various modalities. Subsequently, in \cite{wang2021mixed}, a hybrid 2D and 3D convolutional network with pyramid contexts was introduced. Recently, ALMN \cite{zhou2022three} proposed a multiple subnetworks fusion model via 3D affinity learning to enhance the performance of breast tumor segmentation. The experimental findings illustrate that more intricate networks have the capability of generating superior segmentation performance compared to simpler networks. The second category of methods focuses on designing novel learning frameworks. For instance, in \cite{zhang2018hierarchical}, a mask-guided hierarchical learning framework was implemented for lightweight networks to perform coarse-to-fine segmentation for breast tumors. Recently, Wang $et~al$. \cite{wang2021breast} introduced a synthesis loss to optimize the baseline segmentation networks through a tumor-sensitive synthesis module. Despite early promise, the complex networks or optimization protocols of these models are difficult to generalize across datasets acquired from various medical centers, population and protocols. For example, recent segmentation networks that achieve satisfactory performance, such as ALMN \cite{zhou2022three}, may incur high computation costs. Additionally, the efficacy of mask-guided hierarchical learning \cite{zhang2018hierarchical} or synthesis loss \cite{wang2021breast} may deteriorate when the distributions of MRI images vary. Therefore, it is crucial to design novel network architecture and optimization strategy to achieve a trade-off between overall accuracy and computation costs.\par
\indent In this study, we propose an efficient encoder-decoder based hybrid network with the integration of convolution layers, deconvolution layers and transformer layers for breast tumor segmentation. Recently, the hybrid network (i.e., amalgamation of CNN with transformers) has been demonstrated to be a highly effective strategy to enhance network capacity \cite{chen2021transunet,wang2021transbts,xie2021cotr} for medical semantic segmentation. Particularly, the self-attention mechanism of transformer is able to dynamically adjust the receptive field on the basis of input contents, and is therefore superior to convolutional operations in modeling the long range dependency. Specifically, we design a 3D CNN encoder consisting of only three convolution blocks, followed by 3D transformer layers to capture the global dependency between bottleneck features.
To enhance optimization efficiency, two subnetwork encoders are employed, and three transpose convolution layers are utilized for decoder.
In other words, the first encoder subnetwork is dedicated to extract features for the skip connection to the decoder, while the second one is  designed to extract bottleneck features for the transformer layers. By contrast, the decoder with the mixed feature information from transformer and encoder subnetworks enables the generation of segmentation masks at full resolution.\par
\indent To further improve the segmentation performance of the proposed hybrid network, we introduce a novel prototype learning guided prediction module. During the training process, we calculate prototypical features for each category using efficient online clustering. Once the category-specific prototypical features are obtained, we calculate the similarity maps between these prototypes and the normalized output features of decoder, which can provide localization maps of tumor voxels. In addition, we design an attention-based fusion strategy to reorganize the output features from the decoder. Finally, we fuse the similarity maps with the prototypes and the reorganized output features of the decoder to generate the final breast tumor masks. This approach allows the hybrid network to capture both global and local semantic cues more effectively.\par
To address the issue that excessively long training iterations are required to focus on self-attention in a hybrid network, a two-stage optimization strategy is employed. In the first stage, the encoder subnetworks and decoder in the hybrid network are optimized, followed by the optimization of the prototype guided prediction module and transformer layers in the second stage, using an end-to-end approach. Additionally, a training image sampling strategy is proposed to effectively handle the problem of class-imbalance of breast tumor voxels during the optimization process.\par
\indent Given the pivotal role of Human Epidermal Growth Factor Receptor 2 (HER2) in cell growth and survival pathways, the accurate identification of HER2-positive and HER2-negative cases is essential for guiding treatment decisions and therapy planning \cite{goddard2011her2}. In order to assess the clinical utility of the proposed breast tumor segmentation algorithm in streamlining clinician workload, we have developed a HER2 status diagnosis pipeline utilizing features derived from segmented breast tumors by PLHN. Radiomics features are extracted from both automatically segmented masks and manually-segmented tumor masks for predictive purposes. The findings indicate that the automatic segmentation by PLHN aligns closely with manual segmentation in predicting HER2 status.\par
\indent In conclusion, the main contributions of the paper can be summarized in the following three aspects:\par
\begin{itemize}
\item A prototype learning guided hybrid network (PLHN) is proposed, which incorporates two parallel encoder subnetworks to effectively segment breast tumors by combining CNN with transformer layers.
\item A prototype guided prediction module is designed where representative prototypical features for each category are calculated using online clustering. The similarities between these features and prototypes are fused with the output features of decoder to improve the segmentation performance of breast tumors.
\item A two-stage optimization strategy has been developed to enhance the efficiency of optimizing hybrid network, and a sampling strategy has been designed to address the class imbalance problem.
\item We evaluate the segmentation performance on two breast DCE-MRI datasets, including one large-scale in-house dataset and one public dataset. The proposed method PLHN achieves the best segmentation performance against other SOTA segmentation methods.
\end{itemize}
\section{Related Work}
 In this section, some representative prior methods on learning based breast tumor segmentation in DCE-MRI, hybrid network combining CNN and transformer, and prototype learning guided image segmentation will be reviewed.
\vspace{-0.3cm}
\subsection{Learning based Breast Tumor Segmentation in DCE-MRI}
Deep learning based segmentation frameworks have recently achieved impressive performance in the field of breast MRI. For instance, a mask-guided hierarchical learning framework was proposed in \cite{zhang2018hierarchical}, where two cascaded fully convolutional networks were designed to accurately detect tumor regions with guidance from breast masks. In \cite{lv2021temporal}, the spatial contextual dependency of inter-slice and temporal contextual dependency of inter-sequence were explored. A graph temporal attention module was designed to integrate the temporal-spatial information hidden in MRI images into tumor segmentation. In \cite{wang2021breast}, a novel segmentation network with tumor-sensitive synthesis was designed to achieve accurate breast tumor segmentation, which made full use of the contrast-enhancement
characteristics of breast tumors and designed a synthesis module to suppress false segmentations. In \cite{zhou2022three}, a multi-branch ensemble network was proposed, in which two types of subnetworks were combined for breast tumor segmentation. Moreover, an end-to-end trainable 3D affinity learning based refinement module was designed to refine the segmentation outputs effectively. Recently, a diffusion kinetic model
(DKM) was proposed in \cite{lv2023diffusion} to implicitly exploit hemodynamic priors in DCE-MRI and effectively generate high-quality segmentation maps only requiring images.\par
Nevertheless, these methods have certain limitations that hinder their ability to produce tumor masks with high accuracy. For instance, ALMN exhibits state-of-the-art performance but introduces a two-branch architecture that results in high computation costs. On the other hand, diffusion-based methods show promising segmentation performance but suffer from high inference complexity. In contrast, our approach focuses on the design of a hybrid network that combines lightweight CNN layers with transformer blocks in order to strike a better trade-off between computation costs and segmentation performance.
\vspace{-0.3cm}
\subsection{Hybrid Network for Semantic Segmentation}
The hybrid networks combining CNN and transformers has shown advanced performance in medical field. For instance, in \cite{wang2021transbts},
3D CNN was utilized by the encoder to extract the local 3D spatial feature maps. Meanwhile, the feature maps were reformed elaborately for tokens that were fed into transformer for global feature modeling. In \cite{xie2021cotr}, the CNN was constructed to extract feature representations and an efficient deformable transformer (DeTrans) was built to model the long-range dependency on the extracted feature maps. In \cite{wang2022smeswin}, a superpixel and EA-based Segmentation network (SME Swin-Unet) were proposed from the CNN and ViT mixed-wise perspective to provide the precise and reliable automatic segmentation of medical images. In \cite{luo2022semi}, a cross teaching strategy between CNN and transformer was designed for semi-supervised segmentation, considering the difference in learning paradigm between CNN and transformer rather than CNNs alone. In \cite{lee20223d}, a lightweight volumetric ConvNet, termed 3D UX-Net, was proposed in which the hierarchical transformer was adapted using ConvNet modules for robust volumetric segmentation. Recently, an effective CNN-Transformer framework was proposed in \cite{lin2023lighter}, by introducing novel techniques such as self-supervised attention for speeding up the convergence of transformers, gaussian-prior relative position embedding, and query-wise and dependency-wise pruning.\par
Unlike existing methods, we have developed a simple while effective hybrid segmentation network that employs an encoder-decoder architecture with two parallel encoder subnetworks, enabling the integration of transformer layers into the hybrid network. It is worth noting that a specialized encoder subnetwork is employed to extract bottleneck features for the 3D transformer layers, enabling the capture of global dependency between these features and accelerating the convergence speed of the hybrid network.
\vspace{-0.3cm}
\subsection{Prototype Learning Guided Image Segmentation}
\indent The notations of prototypes have already been used to boost the segmentation performance \cite{zhou2022rethinking,liang2022gmmseg,zhou2022regional,xu2022semi}. For example, in \cite{zhou2022rethinking}, each category of semantic segmentation was represented by a set of non-learnable prototypes corresponding to the mean features of several training pixels per each category. The dense prediction for segmentation was therefore achieved by a non-parametric nearest prototype retrieving approach. In \cite{zhou2022regional}, a novel RCA method which was designed for weakly supervised segmentation using only image-level supervision. In detail, RCA was equipped with a continuously updated memory bank for storing massive historical pseudo-region features. In \cite{xu2022semi}, a prototype-based predictor was integrated into the semi-supervised semantic segmentation network and a novel prototype-based consistency loss was then proposed to regularize the intra-class feature representation to be more compact. Recently, an Uncertainty-informed Prototype Consistency Learning (UPCoL) framework was designed in \cite{lu2023upcol} to fuse prototype representations from labeled and unlabeled data judiciously by incorporating an entropy-based uncertainty mask. In this way, a more discriminative and compact prototype representation can be learned for each class by enforcing consistency constraint on prototypes. To utilize  limited annotated data effectively in semi-supervised segmentation in \cite{zhang2023self}, a self-aware and cross-sample prototypical learning method was designed to enhance the diversity of prediction in consistency learning.\par
\indent Distinct from current solutions which utilize prototypical features from the perspective of contrastive learning or consistency learning, we exploit to integrate prototypes into the hybrid network as important guidance to generate more accurate tumor masks. To the best of our knowledge, this is the first time that prototypes have been utilized to address the challenging task of breast tumor segmentation in DCE-MRI.
\vspace{-0.3cm}
\begin{figure*}
\centering
\centerline{\includegraphics[width=15cm]{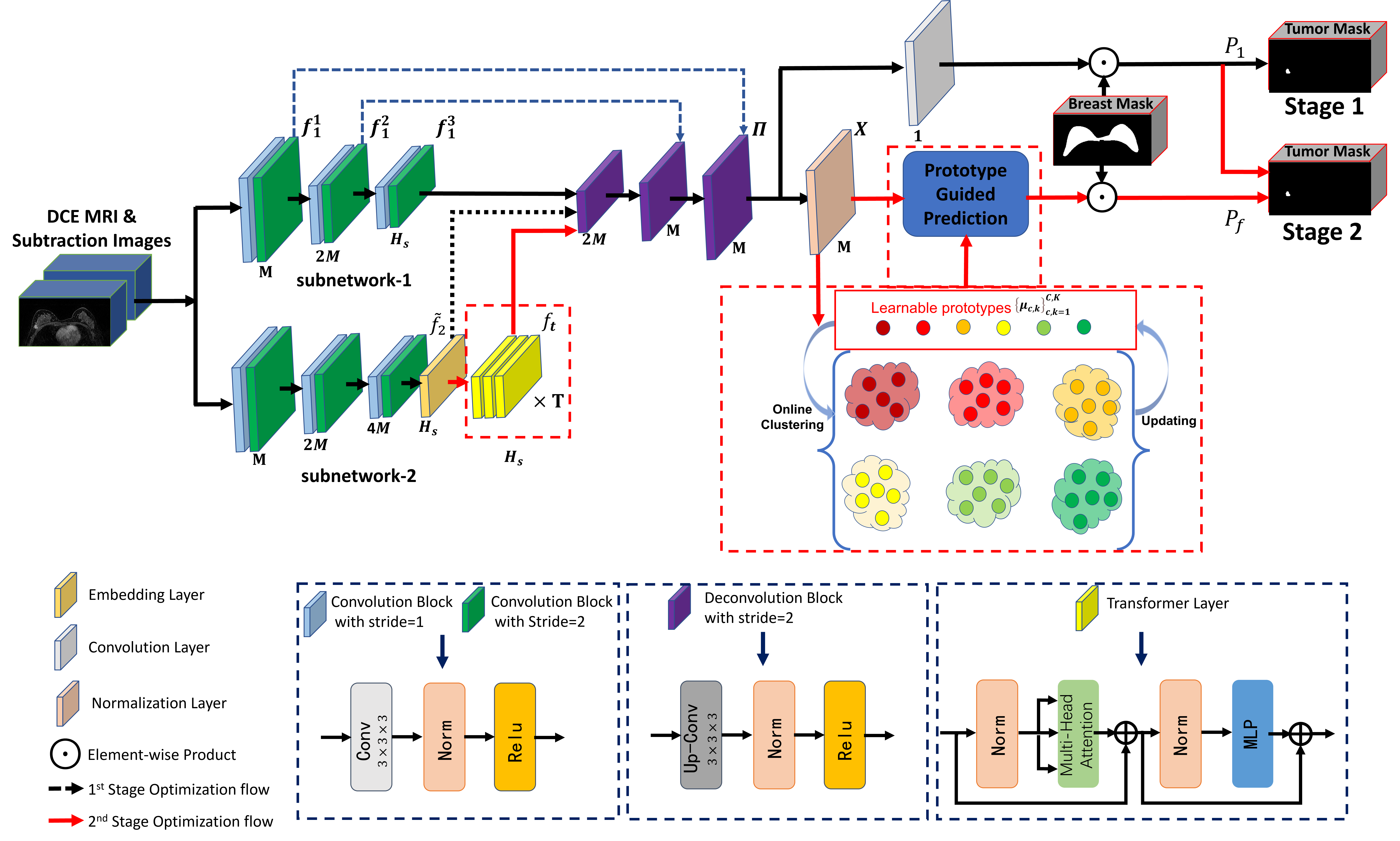}}
\vspace{0.1cm}
\caption{An overview of the proposed prototype learning guided hybrid network. In the first stage of optimization, the concatenation of $[f_1^3,\tilde{f}_2]$ is fed into the decoder and the network is optimized on $P_1$. In the second stage of optimization and inference stage, the concatenation of $[f_1^3,f_t]$ is fed into the decoder and the network is optimized on $P_1$ and $P_f$ jointly.}
\label{fig:fw}
\vspace{-0.4cm}
\end{figure*}
\section{The Proposed Method}
Let $D=\{(X_n,Y_n)\}_{n=1}^N$ be the training dataset consisting of $N$ pairs of input patches $\{X_n\}$ and their corresponding annotations $\{Y_n\}$ for $n=\{1,\cdots,N\}$. The $X_n$ is with size $2 \times H \times W \times Z$ and $Y_n$ is with size $H \times W \times Z$, where $H$, $W$ and $Z$ stand for the height, width, and channel of the input patch. Following the setting in ALMN \cite{zhou2022three}, $X_n$ contains two channels with the first one being the post-contrast image and the second one being the subtraction image between the first post-contrast image and its corresponding pre-contrast image. The learning objective of the proposed method is to produce the segmentation results of breast tumors in DCE-MRI.\par
In the following subsections, the architecture of our proposed hybrid network is first described in subsection \ref{sec:1}. Secondly, the strategy of prototype learning guided breast tumor segmentation is presented in subsection \ref{sec:2}. Thirdly, the overall optimization protocol is described in subsection \ref{sec:3}. Finally, the details of Computer-Aided Diagnosis (CAD) system for breast cancer diagnosis is introduced in subsection \ref{sec:4}.
\subsection{Architecture of Hybrid Network}\label{sec:1}
The overall architecture of the proposed hybrid network is illustrated in Fig. \ref{fig:fw}. A U-shape encoder-decoder architecture with skip connection is designed to build the backbone network. Then, transformer layers are integrated into the backbone network, so as to model the global correlations between voxels.
\subsubsection{Encoder Subnetworks}
The encoder subnetworks are built by stacking 3D convolution blocks. The 3D convolution block is the basic component of this network, which consists of a batch normalization layer \cite{ioffe2015batch}, a Leak ReLU activation function, and a convolutional layer. It is observed that replacing the patchify stem with a simple convolutional stem leads to a remarkable change in optimization behavior, and transformer converges faster with the convolutional stem \cite{xiao2021early}. In our implementation, dense small convolutions are employed to significantly reduce the number of parameters while preserving the receptive field. The encoder subnetwork-1 consists of six 3D convolution blocks whose kernel sizes are $3\times 3\times 3$, and the features will be downsampled by 8 times. Correspondingly, the intermediate features $f^1_1$, $f^2_1$ and $f^3_1$ (as shown in Fig.~\ref{fig:fw}) extracted from encoder subnetwork-1 will be fed into the decoder via skip connection. Specially, $f^3_1$ is of size $h \times w \times z \times H_s$, where $h$=$\frac{H}{8}$, $w$=$\frac{W}{8}$, $z$=$\frac{Z}{8}$ and $H_s$ is the hidden size of following transformer layers.\par
To enhance the optimization efficiency of the hybrid network, a separate encoder subnetwork-2 has been designed specifically for the subsequent transformer layers. The network architecture of encoder subnetwork-2 is of slightly difference with the parameters of encoder subnetwork-1, and its corresponding output feature is indicated as $f_2 \in R^{h \times w \times z \times C_i}$, where $M$ is the configuration parameter of feature dimension and $C_i$=$4M$ represents the channel number of feature map. Then, the bottleneck feature $f_2$ extracted from encoder subnetwork-2 is fed into the following 3D transformer layers, so as to capture the global contextual dependency between features of breast tumors.
\subsubsection{3D Transformer Layer}
A linear embedding layer is first designed, in which the input feature $f_2$ is reshaped to resolution $(h.w.z) \times C_i$ and then it is projected into a $H_s$ dimensional embedding space $\tilde{f}_2 \in R^{(h.w.z)\times H_s}$ . \footnote{$\tilde{f}_2$ will be reshaped to the size $h \times w \times z \times H_s$ if it is fed into decoder directly}
Then, the non-shifted version of 3D Swin-Transformer layers designed in \cite{zhou2023nnformer} are stacked to construct the transformer layers.
In the transformer layer, $\tilde{f}_2$ is first divided into flattened uniform non-overlapping patches $\bar{f_2} \in R^{L \times W_s^3 \times C_i}$, where $W_s$ is the size of local patch and $L=\frac{h \times w \times z}{W_s^3}$ denotes the sequence length. The transformer layer comprises of Multi-head Self-attention (MSA) and multilayer perceptron (MLP) sublayers, which are defined as:\par
\begin{equation}
\begin{aligned}
&\bar{f}_2^{l} =MSA(Norm(\bar{f}_2^{l-1}))+\bar{f}_2^{l-1}, \\
&\bar{f}_2^{l}=MLP(Norm(\bar{f}_2^{l}))+\bar{f}_2^{l},\\
\end{aligned}
\end{equation}
where $l$ is the intermediate block identifier, Norm() represents the operation of normalization. MLP consists of two linear layers with one GELU activation function and two dropout function. Then, $T$ successive transformer layers are utilized to construct the transformer block. Finally, the output feature of transformer layer is defined as $f_t=\bar{f_2}^{T}$ and $f_t$ will be reshaped to the resolution $h \times w \times z \times H_s$.
\subsubsection{Decoder}
As shown in Fig.~\ref{fig:fw}, the network architecture of decoder is highly symmetric to that of encoder subnetworks. The feature $f_t$ will be fed into the decoder, followed by feature upsampling 3D deconvolution layers. Similar to the convolution blocks, the 3D deconvolution block consists of a transpose convolution layer, a batch normalization layer and a Leak ReLU activation function. In detailed, the decoder is composed of three deconvolution blocks. It is noted that the intermediate features $f^1_1$, $f^2_1$ and $f^3_1$ extracted from encoder subnetwork-1 will be combined with the intermediate features extracted from decoder via skip connection, so as to capture the semantic and fine-grained information. For example, the combination of $f^3_1$ and $f_t$ will be fed into the decoder. In detail, the feature extracted from the second last layer of decoder is defined as $\Pi=\{\pi_n \in R^M,n \in \{1,..,H\times W \times Z\}\}$, where $M$ represents the channel number of feature $\Pi$. Moreover, $\Pi$ will be normalized to $X=\{x_n \in R^M,n \in \{1,..,H\times W \times Z\}\}$ for the following operation as prototype guided prediction. Given the feature $\Pi$, the last faceted extension block in decoder also employs a convolution with kernel size $1\times 1 \times 1 \times M \times 1$ to generate the final tumor prediction probability. Finally, the convolution layer for tumor segmentation will transform $\Pi$ to prediction $P_1$ together with the sigmoid function:
\begin{equation}
P_1=sigmoid(conv(\Pi)).
\end{equation}
It is noted that $P_1$ will only serve as the intermediate result in the optimization procedure.
\vspace{-0.3cm}
\subsection{Prototype Guided Breast Tumor Prediction} \label{sec:2}
To address the representation limitation of parameterized neural networks, a novel prototype-guided strategy is proposed for breast tumor segmentation. The prototype refers to representative features or sub-centers associated with each category. In our implementation, the prototypes are used to capture the semantic properties of the respective categories, eliminating the need for additional optimization of learnable parameters.
\subsubsection{Definition of Prototypes}
Following the setting in \cite{zhou2022rethinking}, $K$ non-learnable prototypes are learned for each category and the category number is $C$. More specifically, each class $c \in \{1,...,C\}$, is represented by a total of $K$ prototypes $\{\mu_{c,k}\}_{k=1}^{K}$. In this way, $CK$ prototypes are obtained, i.e, $\{\mu_{c,k} \in R^M\}_{c=1,k=1}^{C,K}$. In the training procedure, the pixels belonging to category $c$ in a training batch is defined as $I^c =\{i_n\}_{n=1}^{N_c}$ according to the ground truth masks, where $N^c$ represents the number of voxels belonging to category $c$ in a batch. The goal of prototype learning is to map the pixels $I^c$ to the $K$ prototypes $\{\mu_{c,k}\}_{k=1}^K$ of category $c$. In the first iteration of the optimization procedure, prototype $\mu_{c,k}$ is determined as the center of $k$-th sub-cluster of training voxel samples belonging to class $c$ in the feature space $X^c=\{x^c_n\}_{n=1}^{N^c}$, where $x^c_n$ stands for the the normalized embedding of voxel $n$. Given the prototypes, the voxel-to-prototype assignment in each batch is defined as a matrix $\Upsilon^c=\{\Upsilon^{c,k}_{n}\}_{n=1}^{N^c} \in \{0,1\}^{CK \times N^c}$, where $\Upsilon^{c,k}_{n}$ is defined as one-hot assignment vector of voxel in over the $CK$ prototypes. In our implementation, $\Upsilon^{c,k}_{n}$ is defined according to the learning based similarities between voxel embedding $X^c$, and the set of prototypes $\{\mu_{c,k}\}_{k=1}^K$. Then the category prediction $\Upsilon^c$ of each voxel $i\in I^c$ is achieved by a winner-take-all classification:
\begin{equation}
\Upsilon^{c^*,k^*}_{n}=1, \text{where} (c^*,k^*)=\text{arg~max}_{(c,k)}\{\Omega(x^c_n,\mu_{c,k})\}_{c,k=1}^{C,K},
\end{equation}
where the function $\Omega()$ represents the distance measurement. For example, the $\Omega$ is defined as the cosine similarity, $\Omega(x_i,\mu)=x_i^{T}\mu$ in \cite{zhou2022rethinking}. In our implementation, the MLP is used to construct a distance measurement network $\Omega$=$D_n$. Then, the concatenated feature $[x_i,\mu]$ will be fed into $D_n$ to produce the similarity measurements in an end to end manner. Finally, the indicator $\Upsilon^{c^*,k^*}_{n}$=1 represents that the feature vector $x_n$ is assigned with prototype label $(c^*,k^*)$, otherwise $\Upsilon^{c^*,k^*}_{n}$=0. Then, the details of how the prototypes will be generated by online clustering is presented in the next subsection.\par
\vspace{-0.1cm}
\subsubsection{Learning Prototypes via Online Clustering}
In the training procedure, the problem of prototype assignement can be formulated as an online clustering problem:
\begin{equation}
J=\sum\limits_{n=1}^{\tilde{N}T_s} \sum\limits_{c=1}^C\sum\limits_{k=1}^K \Upsilon^{c,k}_{n} \parallel x_n^c-\mu_{c,k} \parallel^2,
\end{equation}
where $\Upsilon^{c,k}_{n} \in \{0,1\}$ is the indication variable, $T_s$ is the iteration number and $\tilde{N}$ is the voxel number in a batch for each iteration. The non-learnable prototypes $\{\mu_{c,k}\}_{k=1}^K$ can be treated as the centers of the corresponding embedded voxel samples. It is obvious that the online clustering formulation is scalable to large amounts of data. The key idea for solving the clustering problem defined is that pixels belonging to the same category are assigned to the prototypes calculated for that category. In order to update the prototypes effectively, after each training iteration, each prototype is updated as Robbin-Monro stochastic approximation procedure \cite{dvoretsky1955stochastic}. The prototypes are evolved continuously by accounting for the online clustering results:
\begin{equation}
\mu_{c,k}^{t}=\eta \mu_{c,k}^{t-1}+(1-\eta)R_{c,k},
\end{equation}
where $\eta \in \{0, 1\}$ is a momentum coefficient, and $t \in \{1,..,T_s\}$ is the iteration number for learning prototypes. The $R_{c,k}$ indicates the normalized, mean vector of the embedded training voxels which is define as:
\begin{equation}
R_{c,k}=\frac{\sum\limits_{n=1}^{\tilde{N}} \Upsilon^{c,k}_{n}x^{c}_{n}}{\sum\limits_{n=1}^{\tilde{N}} \Upsilon^{c,k}_n}.
\end{equation}
After $T_s$ iterations, the final prototypes $\mu_{c,k}^{T_s}$ is utilized for the inference of breast tumors.
\subsubsection{Fusing Prototypes for Breast Tumors Segmentation}
\begin{figure}
\centering
\centerline{\includegraphics[width=9cm]{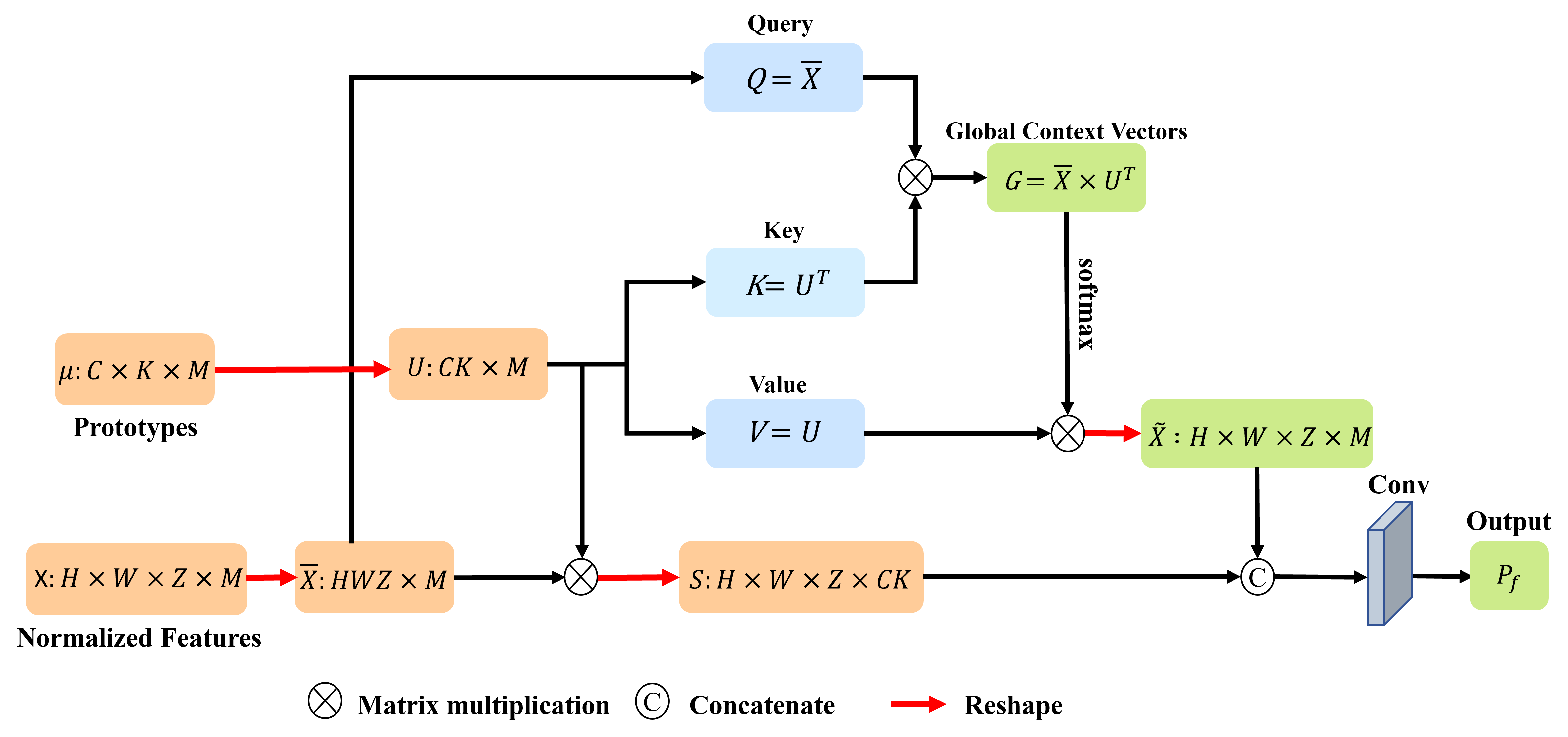}}
\caption{The overall pipeline of prototype guided breast tumor segmentation is illustrated by taking the normalized feature $X$ and the prototype $\mu$ as inputs. The dimension of feature is also listed.}
\label{fig:pred}
\vspace{-0.5cm}
\end{figure}

The overall pipeline of prototype guided breast tumor segmentation is illustrated in Fig. \ref{fig:pred}. Given the normalized feature $X$ and the set of prototypes $\{\mu_{c,k} \in R^M\}_{c,k=1}^{C,K}$, a similar tensor $S=\{s_n\}$ is first defined for voxel $n$:
\begin{equation}
s_n(c,k)=\Omega(x_n,\mu_{c,k}).
\end{equation}
The tensor $S$ is reshaped to the size $H \times W \times Z \times CK$ and the set of prototypes $\{\mu_{c,k} \in R^M\}_{c,k=1}^{C,K}$ can be organized as a matrix $U$ with shape $CK \times M$. For the subsequent operation, $X$ is rearranged to a matrix $\bar{X}$ with shape $HWZ \times M$.\par
In order to extract more discriminative features, a query-key-value ($QKV$) attention module is defined to refine $\bar{X}$ by fusing $\bar{X}$ with $U$:
\begin{equation}
Attention(Q,K,V)=softmax(\frac{QK^T}{\sqrt{d_k}})V,
\end{equation}
where $Q$, $K$, $V$ denote the query, key and value matrices. The $d_k$ is the scaling factor and it is defined as $d_k$=1. We set $Q=\bar{X}$, $V=K=U$. In this way, the transformed feature $\bar{X}$ will be rearranged to $\tilde{X}=Attention(\bar{X},U,U)$, and $\tilde{X}$ is finally reshaped to the size as $H \times W \times Z \times M$.\par
Different from the non-parametric prediction strategy which solely defined on comparing $S$ in \cite{zhou2022rethinking}, the transformed feature $\tilde{X}$ will be concatenated with the similar tensor $S$ to produce the final segmentation probability via convolution:
\begin{equation}
P_f=sigmoid(conv(cat[\tilde{X},S])),
\end{equation}
where the convolution operation $conv$ stands for the kernel size as $3 \times 3 \times 3 \times (M+CK) \times 1$.\par
\indent Following the practice in \cite{zhou2022three}, the final segmentation mask $M_f(i)$ of breast tumors for the $i$-th node can be generated by setting a hard threshold, such as 0.5:
	\begin{equation}
		M_f(i) = \begin{cases}
		1,& if~P_f(i)\geqslant 0.5; \\
		0,& if~P_f(i) \lessdot 0.5,
		\end{cases}
	\end{equation}
where 1 represents breast tumor mask and 0 is background.
\vspace{-0.3cm}
\subsection{Two-stage Optimization Strategy}\label{sec:3}
\subsubsection{Loss Functions}
\indent The details for loss functions will then be described. During the training process, the Dice loss $L_{Dice}$ is defined as:
\begin{equation}
 L_{Dice}(P,Y)=1-\frac{2\sum\limits_{i=1}^{\bar{N}} P_iY_i}{\sum\limits_{i=1}^{\bar{N}} P_i^2+\sum\limits_{i=1}^{\bar{N}} Y_i^2},
\end{equation}
 where $\bar{N}$ is the voxel number of the input patches, $P_i \in [0.0, 1.0]$ represents the voxel value of the predicted probabilities and $Y_i \in  \{0, 1\}$ is the voxel value of the binary ground truth volume. Correspondingly, the cross-entropy loss $L_{bce}$ is formulated as:
\begin{equation}
 L_{bce}(P,Y)=\sum\limits_{i=1}^{\bar{N}} Y_i log(P_i)+\sum\limits_{i=1}^{\bar{N}} (1-Y_i)log(1-P_i).
\end{equation}
In the training procedure, the combined loss is defined as:\\
\begin{equation}
Lc(P,Y)=L_{Dice}(P,Y)+L_{bce}(P,Y).
\end{equation}
\indent In order to produce more discriminative prototypical features, the voxel-prototype contrastive loss defined in \cite{zhou2022rethinking} is applied as well. After all the samples in current batch are processed, each voxel $i$ is assigned with optimal category index $c^*$ and prototype index $k^*$ according to Eq. (3). The voxel-prototype contrastive loss is utilized to maximize the prototype assignment posterior probability, which is defined as:\\
\begin{small}
\begin{equation}
\ell_{ppc}(i)=-\log \frac{exp(\Omega(x_i,\mu_{c^*,k^*})/\tau)}{exp(\Omega(x_i,\mu_{c^*,k^*})/\tau)+\sum\limits_{\mu^{-} \in U^{-}} exp(\Omega(x_i,\mu^{-})/\tau)},
\end{equation}
\end{small}
where $\Omega$ defines the distance measurement, $x_i$ is the normalized feature for voxel $i$, $U^{-}=\{\mu_{c,k}\}_{c,k=1}^{C,K}/\mu_{c^*,k^*}$, and the temperature $\tau$ controls the concentration level of representations.\par
Two segmentation outputs $P_1$ defined in Eq. (2) and $P_f$ defined in Eq. (9) are used to supervise the optimization procedure. Finally, the overall loss is defined as:
\begin{equation}
\ell_{all}=Lc(P_1,Y)+\lambda_1 \times Lc(P_f,Y)+\lambda_2 \times \ell_{ppc},
\end{equation}
where the weights $\lambda_1$ and $\lambda_2$ changes in the optimization procedure.
\subsubsection{Patch Sampling Strategy}
Another issue that arises pertains to the imbalanced distribution of randomly sampled patches. This is primarily due to the breast tumors occupying only a small fraction of the breast regions, resulting in random patch sampling strategies tending to extract samples dominated by background voxels.
To generate more balanced training samples, a straightforward patch sampling strategy is devised. During the sampling procedure, three patches are extracted in each iteration, with one patch encompassing the entire tumor regions and the remaining two patches containing partial tumor regions.
Then, $3\times B$ patches are extracted in each batch. This enables the sampling of more balanced training patches.\par
\subsubsection{Optimization Strategy}
\indent To address the challenge of optimizing hybrid networks with transformers that require more iterations, a two-stage optimization strategy is proposed.\par
\textbf{Stage 1:} In contrast to the approach of simultaneously optimizing all network parameters at the outset, the two encoder subnetworks' parameters and the decoder of the hybrid network undergo iterative updates across 300 epochs during the initial phase. As shown in the network architecture of Fig. \ref{fig:fw}, the concatenation of $[f_1^3,\tilde{f}_2]$ is inputted into the decoder, while the transformer layers and prototype guided prediction module remain inactive. Ultimately, the corresponding loss function is solely defined on the segmentation output $P_1$. Consequently, we assign values of $\lambda_1$=0 and $\lambda_2$=0 in Eq. (15).\par
\textbf{Stage 2:} Subsequently, in the second phase, the network undergoes joint optimization for an additional 200 epochs with a reduced initial learning rate, incorporating transformer layers and a prototype guided prediction module. The concatenation of $[f_1^3,f_t]$ is input into the decoder, and the loss function is defined on the segmentation outputs $P_1$ and $P_f$ collectively. Additionally, in Eq. (15), the values of $\lambda_1$ and $\lambda_2$ are set to 0.2 and 0.05, respectively, based on the cross-validation experiment in Table \ref{t:wei}. This approach aims to ensure that the optimization procedure for the hybrid network converges within a shorter training period.\par
\vspace{-0.3cm}
\subsection{CAD System for Breast Cancer Diagnosis}\label{sec:4}
In the process of diagnosis, radiomics features are derived from regions of interest (ROIs) through an in-house feature analysis program implemented in Pyradiomics (http://pyradiomics.readthedocs.io) \cite{van2017computational}. These features can be classified into three main groups: (1) geometry features, which characterize the three-dimensional shape attributes of the ROIs; (2) intensity features, which portray the statistical distribution of voxel intensities within the ROIs; and (3) texture features, which delineate the patterns or spatial distributions of intensities at second- and high-orders. To extract texture features, a variety of methods are utilized, including the gray-level co-occurrence matrix (GLCM), gray-level run length matrix (GLRLM), gray-level size zone matrix (GLSZM), and gray-level dependence matrix (GLDM). In our approach, a manually crafted feature vector with a dimension of 1648 is obtained for each modality (pre-contrast or post-contrast). This feature vector comprises first-order features (342 dimensions), shape features (14 dimensions), GLRLM features (304 dimensions), GLSZM features (304 dimensions), GLDM features (266 dimensions), and GLCM features (418 dimensions). The features from both modalities are merged, resulting in a total of 3296 radiomics features for subsequent classification. Following feature extraction, Mann-Whitney U test, Pearson test, and LASSO regression are sequentially employed to reduce the feature dimensionality. Ultimately, a machine learning classifier, specifically logistic regression \cite{demaris1995tutorial}, is applied to the selected features to predict the HER2 status.
\vspace{-0.3cm}
\section{Experimental Results}
\subsection{Dataset and Implementation}
\textbf{Internal Dataset for Breast Tumor Segmentation}: The proposed prototype learning guided hybrid network (PLHN) is evaluated on a large-scale DCE-MRI breast tumor dataset, which are acquired from 404 patients diagnosed with breast cancer from Ruijin hospital (RJ-hospital) and 661 cases from Guangdong provincial people's hospital (GD-hospital). DCE-MRI data comprising two phases (1 pre-contrast image and the first post-contrast image) are collected for the experiment of breast tumor segmentation. The breast tumors are meticulously delineated and reviewed by two senior radiologists using voxel-wise labels at each center. The manual delineation results, after careful correction, are considered as the ground truth. The 1065 cases were randomly divided into three groups: 693 cases (65$\%$) for the training set, 106 cases (10$\%$) for the validation set, and 266 cases (25$\%$) for the internal test dataset. The inter-slice resolution of images collected from GD-hospital ranges from 0.44 (mm) to 0.98 (mm), and the slice thickness ranges from 0.5 (mm) to 1.0 (mm). The inter-slice resolution of images collected from RJ-hospital ranges from 0.7 (mm) to 1 (mm), and the slice thickness is 1.5 (mm). During the training and inference stages, all MRI images, together with their corresponding ground-truth, will be resampled to a consistent resolution of $1 \times 1 \times 1~mm^3$ for the purpose of spatial normalization.\par
\textbf{External Testing Set for Breast Tumor Segmentation}: The external dataset comprises 100 public cases sourced from Yunnan Cancer Hospital \footnote{https://zenodo.org/records/8068383}, each encompassing DCE-MRI images featuring six phases, including one pre-contrast image and five post-contrast images. Furthermore, annotations for breast tumors and the entire breast are included. For our study, we have chosen the pre-contrast and initial post-contrast images showing peak contrast for processing, following the same procedures as with the internal dataset.\par
\textbf{Dataset for Thymoma Segmentation}: In order to highlight the robustness and generalizability of our proposed PLHN in a more comprehensive manner, we conducted additional evaluations on diverse and challenging tasks, including thymoma segmentation from CT scans. While our primary focus is on the development of an efficient network architecture, we exclusively utilized CT images to demonstrate the superior performance of PLHN without leveraging information from multiple modalities. The dataset for thymoma segmentation comprised 284 cases obtained from 284 patients with ages ranging from 16 to 83 years, sourced from Shanghai General Hospital. These cases were randomly distributed into three subsets: 213 cases (75$\%$) for training, 28 cases (10$\%$) for validation, and 43 cases (15$\%$) for testing purposes.\par
\textbf{BraTS dataset for Brain Tumor Segmentation}: The BraTS dataset \cite{menze2014multimodal} encompasses a collection of 484 MRI images, each composed of four distinct modalities: FLAIR, T1-weighted (T1w), T1-weighted with gadolinium contrast enhancement (T1gd), and T2-weighted (T2w). This dataset has been divided into three subsets according to an 80:5:15 ratio according to the setting in \cite{shaker2024unetr++}, designated for training, validation, and testing purposes, respectively. The segmentation targets encompass the entire tumor (WT), the enhancing tumor region (ET), and the tumor core (TC).\par
\textbf{Implementation Details}: The proposed hybrid network is implemented using Pytorch \cite{paszke2019pytorch}, and the corresponding experiments are conducted on a Nvidia GeForce GTX V100 GPU with 32G memory. The network weights are initialized by the Xavier algorithm and weight decay is set to $10^{-4}$, the initial learning rate is 0.01. In the training stage, the patches are sampled from the input MRI images with size $128\times128 \times 48$. In the testing stage, the sliding windows are used to go through the whole MRI image with the given stride, such as $64 \times 64 \times 8$. The averaged predictions of the overlapped patches with a constant stride are used to produce the final segmentation. Moreover, the whole-breast segmentation based refined strategy \cite{zhang2023robust} is also utilized to eliminate the over-segmentation outside the breast region. In the optimization procedure for the compared networks, overall 500 epoches will be run with initial learning rate as 0.01 for fair comparison. As to the two-stage optimization strategy for network with transformer (such as PLHN), the first stage will run for 300 epoches with initial lr=0.01, and the second stage will last for another 200 epoches with the smaller initial lr=0.001. The weights of losses are set empirically as: $\lambda_1$=0.2 and $\lambda_2$=0.05 according to cross-validation. To evaluate the proposed method comprehensively, we utilize four metrics, i.e, Dice Similarity Coefficient (DSC), Positive Prediction Value (PPV) and Sensitivity (SEN) to measure the agreement between manually and automatically segmented label maps, and Average Surface Distance (ASD) to measure the average distances between the surfaces of manually and automatically segmented label maps. The 95$\%$ Hausdorff Distance (HD95) metric is utilized for assessing the performance on the BraTS dataset.\par
\vspace{-0.3cm}
\subsection{Comparison with State-of-the-Art Methods}
\begin{table*}[t]
\small
\renewcommand{\arraystretch}{1}
\caption{Segmentation performance achieved by different methods in terms of DSC($\%$), PPV($\%$), SEN($\%$) and ASD(mm) with 95$\%$ confidence intervals on the internal and external test dataset for breast tumor segmentation. $\uparrow$ means the higher value the better and $\downarrow$ means the lower value the better.}
\label{t:t2}
\centering
\setlength{\tabcolsep}{1.3mm}{
\begin{tabular}{c|cccc|cccc}
\toprule[1.2pt]
&\multicolumn{4}{|c|}{\textbf{Internal testing set}} &\multicolumn{4}{|c}{\textbf{External testing set}}\\
\hline
Method &DSC (\%) $\uparrow$  &PPV (\%) $\uparrow$  &SEN (\%) $\uparrow$  &ASD (mm) $\downarrow$ &DSC (\%) $\uparrow$  &PPV (\%) $\uparrow$  &SEN (\%) $\uparrow$  &ASD (mm) $\downarrow$\\
\hline
Vnet \cite{milletari2016v}  &76.7$\pm$2.9 &81.7$\pm$2.8 &79.1$\pm$2.9 &13.4$\pm$5.1 &75.2$\pm$2.3 &79.6$\pm$2.3 &76.8$\pm$2.5 &11.6$\pm$3.3\\
ResUnet \cite{yu2017volumetric} &78.0$\pm$2.7 &81.7$\pm$2.6 & 80.9$\pm$2.7 & 10.3$\pm$4.2 &76.5$\pm$2.1 &\textbf{83.2$\pm$1.9} & 76.6$\pm$2.4 &9.1$\pm$3.0\\
DMFNet  \cite{chen20193d} &76.7$\pm$2.9 &81.8$\pm$2.6 &78.4$\pm$3.0 & 9.9$\pm$3.3 &75.4$\pm$2.3 &77.3$\pm$2.4 &79.8$\pm$2.4 &11.8$\pm$3.2\\
MHL \cite{zhang2018hierarchical}  &77.1$\pm$2.7 &82.4$\pm$2.6 &78.5$\pm$2.7 &8.9$\pm$3.3 &74.7$\pm$2.2 &81.0$\pm$2.4 &74.2$\pm$2.5 &9.8$\pm$3.1 \\
nnUnet \cite{isensee2021nnu} &78.0$\pm$2.4 &75.1$\pm$2.8 &\textbf{87.8}$\pm$1.9 &15.6$\pm$3.0 &74.1$\pm$2.4 &71.2$\pm$2.8 &\textbf{85.1$\pm$1.8} &19.5$\pm$2.9 \\
SwinUnet \cite{cao2022swin} &71.3$\pm$2.7 &75.6$\pm$3.0 &74.5$\pm$2.6 &14.2$\pm$3.7 &66.6$\pm$2.4 &71.9$\pm$3.0 &69.7$\pm$2.5 &16.2$\pm$3.2 \\
TransBTS \cite{wang2021transbts} &78.0$\pm$2.5 &78.9$\pm$2.6 &82.7$\pm$2.4 &8.3$\pm$2.0 &74.4$\pm$2.2 &74.9$\pm$2.7 &80.0$\pm$2.1 &14.0$\pm$2.5\\
MTLN \cite{zhang20213d} &77.6$\pm$2.5 &80.4$\pm$2.4 & 80.3$\pm$2.4 &9.7$\pm$3.7 &77.3$\pm$2.2 &81.3$\pm$2.3 & 77.7$\pm$2.3 &9.3$\pm$3.1\\
UNETR \cite{hatamizadeh2022unetr} &71.2$\pm$2.8 &70.2$\pm$3.2 &81.6$\pm$2.5 &15.5$\pm$3.1 &71.9$\pm$2.4 &73.6$\pm$2.7 &77.4$\pm$2.5 &12.8$\pm$2.1\\
Tumorsen \cite{wang2021breast} &78.1$\pm$2.5 &81.7$\pm$2.4 &80.8$\pm$2.5 &10.1$\pm$3.8 &77.1$\pm$2.1 &80.8$\pm$2.1 &79.3$\pm$2.2 &10.0$\pm$3.0\\
ALMN \cite{zhou2022three} &79.3$\pm$2.5 &\textbf{83.5}$\pm$2.2 & 81.1$\pm$2.6 &8.6$\pm$3.6 &77.7$\pm$2.2 &82.4$\pm$2.2 &77.8$\pm$2.3 &9.0$\pm$3.0\\
UXNET \cite{lee20223d} &77.3$\pm$2.6 &78.9$\pm$2.7 & 81.4$\pm$2.6 & 10.1$\pm$2.5 &73.8$\pm$2.5 &80.0$\pm$2.5 & 73.7$\pm$2.6 & 11.4$\pm$2.4\\
UNETR++ \cite{shaker2024unetr++} &79.6$\pm$2.6 &81.5$\pm$2.6 & 84.1$\pm$2.6 & 9.6$\pm$3.9 &76.6$\pm$2.4 &75.7$\pm$2.8 &82.6$\pm$2.0 &10.4$\pm$3.0\\
\hline
\textbf{PLHN (Ours)} &\textbf{80.6}$\pm$2.3 &82.7$\pm$2.2 &83.3$\pm$2.3 &\textbf{7.3}$\pm$2.9 &\textbf{78.1$\pm$2.1} &81.5$\pm$2.3 &79.7$\pm$2.1 & \textbf{8.9$\pm$2.5} \\
\hline
\toprule[1pt]
\end{tabular}}
\vspace{-0.3cm}
\end{table*}
\subsubsection{Comparison with SOTA Segmentation Methods for Breast Tumor Segmentation}
As for the task of breast tumor segmentation, 13 SOTA 3D medical image segmentation methods: (1) ResUnet \cite{yu2017volumetric}, (2) Vnet\cite{milletari2016v}, (3) MHL \cite{zhang2018hierarchical}, (4) MTLN \cite{zhang20213d}, (5) DMFNet  \cite{chen20193d}, (6) TransBTS\cite{wang2021transbts}, (7) UNETR \cite{hatamizadeh2022unetr}, (8) UXNET \cite{lee20223d}, (9) ALMN \cite{zhou2022three}, (10) Tumorsen \cite{chen20193d}, (11) nnUnet \cite{isensee2021nnu}, (12) SwinUnet \cite{cao2022swin} and (13) UNETR++ \cite{shaker2024unetr++} are implemented and compared with the proposed \textbf{PLHN}. The corresponding segmentation performance on 266 images of the test dataset is listed in Table \ref{t:t2}. It is evident that the PLHN model outperforms the other 13 SOTA methods when considering metrics such as DSC, PPV, SEN, and ASD in terms of segmentation performance. In comparison to three recently developed methods, namely MHL, ALMN, and Tumorsen, which are specifically tailored for breast tumor segmentation, PLHN exhibits superior performance. Specifically, ALMN achieves the second highest DSC at 79.3$\%$. Conversely, Tumorsen only achieves a DSC of 78.1$\%$, which is nearly equivalent to the baseline ResUnet (78.0$\%$). This observation suggests that the proposed tumor synthesis module does not effectively improve the performance on the current large-scale DCE-MRI dataset, which comprises images collected from various centers. Notably, the lightest network among the compared methods, MHL, only achieves a mean DSC of 77.1$\%$. In contrast, PLHN achieves the highest mean DSC value of 80.6$\%$ with a reasonable increase in parameters. To summarize, PLHN demonstrates a significant improvement of approximately 1.3$\%$ in DSC compared to the current best method, ALMN, which is noteworthy for the advancement of breast tumor segmentation. Additionally, the computational cost of PLHN is 17 times lower than that of ALMN.\par
\indent The proposed PLHN is also compared with other baseline networks which exploit the combination of CNN and transformer, such as TransBTS \cite{wang2021transbts} and UXNET \cite{lee20223d}. Compared with TransBTS \cite{wang2021transbts}, PLHN can be optimized with higher efficiency. For example, the transBTS only achieves a DSC as 76.8$\%$ within 500 epoches when trained from scratch. On the contrary, PLHN without prototypes guided optimization can exhibit a DSC as 78.9$\%$ within 300 epoches. Once the transformer blocks are optimized separately, the DSC of transBTS can be boosted to 78.0$\%$ within 500 epoches. However, it is far lower than 80.6$\%$ of the proposed PLHN which is also obtained within 500 epoches. When compared with one recent method UXNET \cite{lee20223d} which modernizes hierarchical transformer, PLHN demonstrates around 3.3$\%$ DSC improvement (80.6$\%$ vs 77.3$\%$). It is surprising that the transformer-based architecture UNETR achieves a low DSC of 71.2$\%$ and its improved version UNETR++ produces better DSC as 79.6$\%$, suggesting that the proposed hybrid network is better suited for the task of breast tumor segmentation. The experimental results demonstrate that the proposed hybrid network PLHN suits the task of breast tumor segmentation better, with more effective network architecture. To establish the statistical significance of the PLHN model's performance, we executes a two-sample t-test to compare its mean outcomes with those of the leading method, UNETR++, in terms of the DSC and ASD metrics. Importantly, the PLHN model yields average p-values consistently below the 0.05 threshold for both DSC (0.043) and ASD (0.046), signifying that these differences are statistically significant. These findings provide compelling evidence that the PLHN model represents a significant advancement over the prior state-of-the-art method, UNETR++, in the context of breast tumor segmentation.\par
\begin{figure*}[t]
\centering
\begin{overpic}[width=\linewidth]{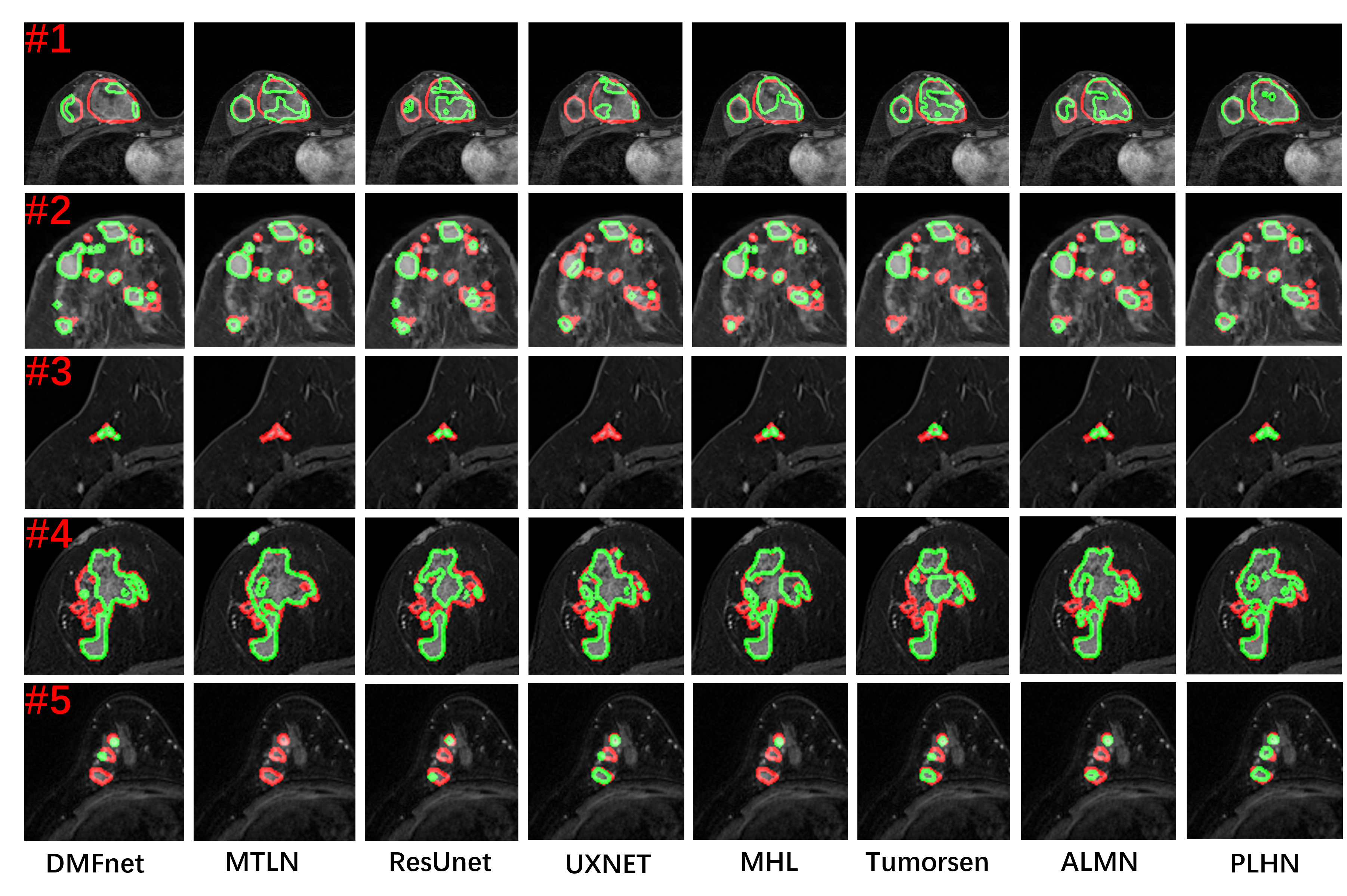}
\end{overpic}
\vspace{-0.5cm}
\caption{The visual comparison of segmentation results between different methods, such as DMFnet, MTLN, ResUnet, UXNET, MHL, Tumrosen, ALMN and PLHN, is displayed. Each row corresponds to one subject, and post-contrast images in axial plane overlaid with ground truth (red line) and automatic segmentation results (green line) of different methods are provided.}
\label{fig:ill}
\vspace{-0.5cm}
\end{figure*}
\begin{figure*}
\centering
\begin{overpic}[width=17cm]{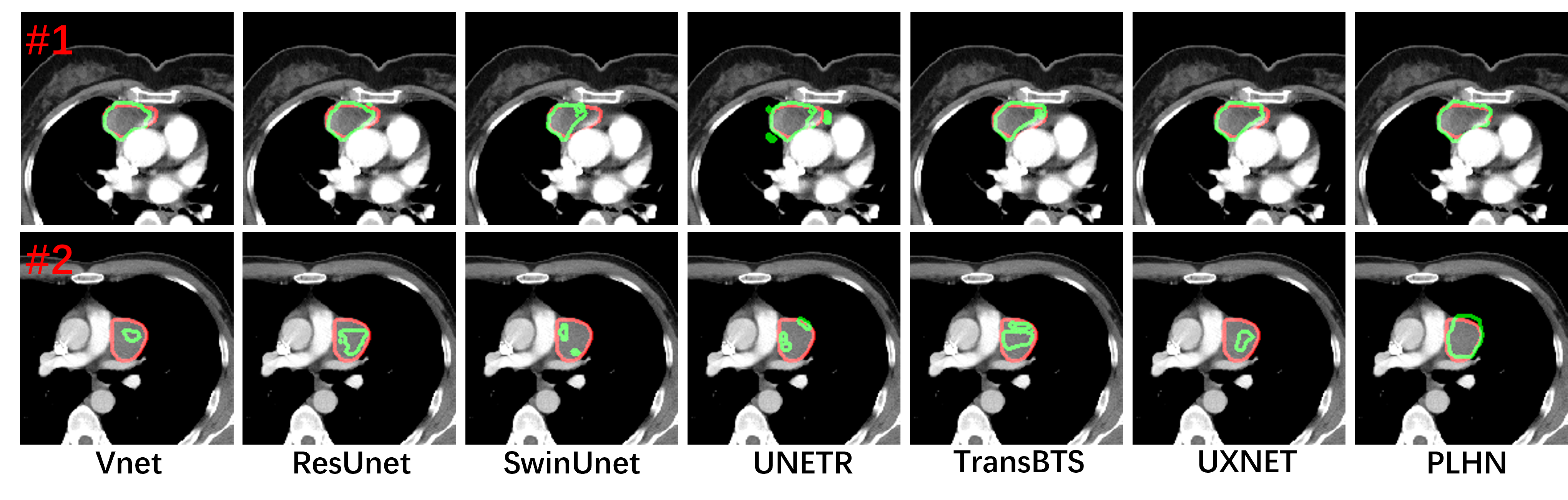}
\end{overpic}
\vspace{-0.5cm}
\caption{The visual comparison of segmentation results between different methods for thymoma segmentation, such as Vnet, ResUnet, SwinUnet, UNETR, TransBTS, UXNET and the proposed PLHN, is displayed. }
\label{fig:xiongxian}
\vspace{-0.5cm}
\end{figure*}
\subsubsection{Comparison with SOTA Segmentation Methods on External Breast Segmentation Testing Set}
In order to further illustrate the universality and resilience of our PLHN framework, we have employed a publicly available external testing dataset to assess the segmentation efficacy. The segmentation models, which have been fine-tuned on the internal training dataset, are directly applied to segment all 100 cases. Analogous to our experiments conducted on the internal dataset, we have juxtaposed our PLHN framework with the aforementioned state-of-the-art segmentation models. The outcomes documented in Table \ref{t:t2} pertaining to the external test dataset indicate that the PLHN model achieves the superior DSC of 78.1$\%$, outperforming the contemporary SOTA approach, UNETR++, which obtains a DSC of 76.6$\%$. Consistent with the observations made in a prior study by Zhang et al. \cite{zhang2023robust}, the segmentation performance on the external testing dataset may exhibit a decline. Conversely, the robustness of PLHN is substantiated by its consistent performance on the external testing dataset, surpassing all comparative methodologies.\par
Two subjects (1 and 2) are selected from external test set and three subjects (3, 4 and 5) are selected from internal test set for visual comparison. The visual results depicted in Fig. \ref{fig:ill} suggest that the proposed PLHN outperforms the compared networks, such as DMFnet, MTLN, ResUnet, UXNET, MHL, Tumrosen, and ALMN, in terms of segmenting MASS, small-sized, or non-mass enhancement (NME) tumors. It can be observed that the 3D rendering of segmentation masks produced by PLHN closely resembles the 3D rendering of the manually annotated ground truth tumor masks. Additionally, it has been discovered that while all the compared methods exhibit similar performance in segmenting MASS tumors, they may neglect numerous voxels belonging to tumors or introduce false positives when handling small-sized (subject 1 and 3) or NME tumors (referred to as subjects 2, 4 and 5). On the contrary, PLHN is capable of identifying more voxels associated with small-sized or NME tumors.\par
\subsubsection{Comparison with SOTA Segmentation Methods for Thymoma Segmentation}
The current study extensively compared the proposed method with several SOTA segmentation approaches using the thymoma segmentation dataset. The results, as detailed in Table \ref{t:XIAN}, underscore the robustness and versatility of our method. This task is particularly challenging due to the limited availability of thymoma segmentation in CT images. Specifically, transformer-based or hybrid networks such as TransBTS, SwinUnet, UNETR, UXNET, and UNETR++ were chosen for comparison. Our proposed PLHN achieves the highest DSC at 76.0$\%$, demonstrating the superiority of our hybrid network for thymoma segmentation. In contrast, the other transformer or hybrid networks yields unsatisfactory results. For example, the representative method TransBTS only achieves a DSC, 70.9$\%$, and the recent SOTA UNETR++ obtains a DSC of 75.0$\%$. Several representative examples are shown in Fig. \ref{fig:xiongxian}, where the red contours represent actual tumor annotations and the green contours depict the segmentation outputs from each method. It is evident that our PLHN framework exhibits superior performance, generating segmentation results that closely match the ground-truth tumor annotations. As depicted in Table \ref{t:XIAN}, UNETR++ emerges as the superior baseline method. The average p-value for the DSC comparison between PLHN and UNETR++ is below the 0.05 threshold (0.041), indicating a statistically significant difference. Meanwhile, the average p-value for the ASD is 0.075, which is above the conventional significance level. This suggests that when DSC is considered as the primary evaluation criterion, PLHN is the preferred choice for thymoma segmentation.\par
\begin{table}\small 
\newcommand{\tabincell}[2]{\begin{tabular}{@{}#1@{}}#2\end{tabular}}
\centering
\renewcommand\arraystretch{1}
\setlength{\tabcolsep}{0.7mm}
\caption{The segmentation performance achieved by various methods for thymoma segmentation is evaluated in terms of DSC($\%$), PPV($\%$), SEN($\%$) and ASD(mm) with 95$\%$ confidence intervals on the thymoma segmentation dataset.}
\begin{tabular}{c|cccc}
\toprule[1.2pt]
Method & DSC (\%) $\uparrow$  & PPV (\%) $\uparrow$  &SEN (\%) $\uparrow$  & ASD (mm) $\downarrow$ \\
\hline
Vnet \cite{milletari2016v} &73.5$\pm$6.5 &81.3$\pm$6.3 & 71.0$\pm$7.4 &2.9$\pm$2.1 \\
ResUnet \cite{yu2017volumetric} &68.6$\pm$7.3 &69.4$\pm$7.3 & 76.1$\pm$8.5 &4.8$\pm$3.6 \\
nnUnet \cite{isensee2021nnu} &72.7$\pm$7.9 &73.9$\pm$8.2 &\textbf{80.3$\pm$5.6} &3.3$\pm$2.5 \\
TransBTS \cite{wang2021transbts} &70.9$\pm$7.8 &83.2$\pm$7.4 & 65.9$\pm$8.3 &2.9$\pm$2.3 \\
SwinUnet \cite{cao2022swin} &69.3$\pm$8.3 &75.9$\pm$7.2 & 70.3$\pm$8.8 &6.9$\pm$8.5 \\
UNETR \cite{hatamizadeh2022unetr} &68.4$\pm$7.6 &79.4$\pm$7.4 & 68.6$\pm$7.7 &8.6$\pm$7.9 \\
UXNET \cite{lee20223d} &69.1$\pm$6.9 &69.1$\pm$8.1 &78.7$\pm$5.6 &11.7$\pm$6.5 \\
UNETR++ \cite{shaker2024unetr++} &75.0$\pm$6.3 &84.9$\pm$5.7 &70.8$\pm$7.0 &3.1$\pm$2.6 \\
\hline
\textbf{PLHN (Ours)} &\textbf{76.0$\pm$5.1} &\textbf{86.3$\pm$3.7} &72.6$\pm$6.1 &\textbf{2.6$\pm$2.8}\\
\hline
\toprule[1pt]
\end{tabular}
\label{t:XIAN}
\vspace{-0.2cm}
\end{table}
\subsubsection{Comparison with SOTA Segmentation Methods on BraTS Dataset}
In this experiment, the network architecture of PLHN is incorporated into the training and inference framework of UNETR++. The quantitative results of the experiment conducted on the BraTS dataset are presented in Table \ref{t:brast}. According to the evaluation criteria, PLHN achieves the best performance for the DSC of WT (91.56$\%$) and average HD95 metrics (4.89mm) when compared with other methods such as UNETR++, nnFormer, nnUnet, SwinUnet, TransBTS, and UNETR \footnote{The performance metrics for UNETR++ are generated using the open-source model weights. The metrics of other methods are cited from \cite{shaker2024unetr++} .}. However, UNETR++ yields better DSC metrics for subregions ET and TC, as well as the best average DSC metric (82.68$\%$). To further enhance the segmentation performance, the proposed prototype-guided prediction module is also integrated with UNETR++ to create a new network, PLUNETR++. This combination leads to increased DSC metrics for all subregions, with the best average DSC reaching 82.89$\%$. The statistical significance of PLUNETR++ over UNETR++ is computed by the average p-values for the evaluation metrics: average DSC ($<$0.05) and average HD95 ($<$0.05). The superior performance confirms the effectiveness of the prototype-guided prediction strategy for the task of brain tumor segmentation.
\begin{table}[t]
\newcommand{\tabincell}[2]{\begin{tabular}{@{}#1@{}}#2\end{tabular}}
\centering
\renewcommand\arraystretch{1}
\setlength{\tabcolsep}{1.1mm}
\caption{The segmentation performance of various methods is assessed based on DSC ($\%$) and HD95 (mm) metrics using the brain tumor segmentation test dataset. AVG represents the average value. PLUNETR++ is the integration of the prototypes guided prediction module with UNETR++.}
    \begin{tabular}{l|ccc|c|c}
    \toprule[1.2pt]
       Method &\multicolumn{4}{c|}{DSC ($\%$)~$\uparrow$} & \multicolumn{1}{c}{HD95(mm)~$\downarrow$} \\
       \cline{2-6}
        ~ & WT & ET & TC & AVG &AVG \\
        \midrule
        UNETR~\cite{hatamizadeh2022unetr} & 90.35 & 76.30 & 77.02 & 81.22 &6.61 \\
        TransBTS~\cite{wang2021transbts} & 90.91 & 77.86 & 76.10 & 81.62 &5.80 \\
        SwinUnet~\cite{cao2022swin} & 91.12 & 77.65 & 78.41 & 82.39 &5.33\\
        nnUnet~\cite{isensee2021nnu} & 91.21 & 77.96 & 78.05 & 82.41 &5.78\\
        nnFormer~\cite{zhou2023nnformer} & 91.23 & 77.84 & 77.91 & 82.34 &5.18\\
        UNETR++~\cite{shaker2024unetr++} & 91.16 & 78.46 & 78.43 & 82.68 &5.27\\
        \midrule
        PLHN & \textbf{91.56} & 77.88 & 78.37 & 82.60 &\textbf{4.89} \\
        PLUNETR++ & 91.37 & \textbf{78.64}& \textbf{78.67} & \textbf{82.89} &5.04\\
        \bottomrule
    \end{tabular}
\label{t:brast}
\vspace{-0.2cm}
\end{table}
\subsubsection{Comparison with SOTA Contrastive Learning based Segmentation Methods}
\indent Two recent methodologies, specifically UPCoL \cite{lu2023upcol} and IDEAL \cite{basak2023ideal}, both rooted in contrastive learning principles, have been implemented to enhance the efficiency of the hybrid network. The amalgamation of the prototype-guided contrastive learning technique from UPCoL into the hybrid network yielded a commendable DSC score of 80.1$\%$ as depicted in Table \ref{t:con}. Conversely, the incorporation of the contrastive loss as detailed in \cite{basak2023ideal} led to a lower DSC value of 79.1$\%$, which falls below the 79.4$\%$ achieved without the application of the contrastive learning loss. This observation suggests that the intricately devised prototype-based contrastive learning mechanism \cite{lu2023upcol} is better suited for our specific task when compared to the simplistic metric learning guided contrast approach \cite{basak2023ideal}. In contrast, PLHN demonstrated superior performance by attaining a DSC score of 80.6$\%$ through the utilization of prototypes as guiding elements for the prediction of segmentation masks. Upon the removal of the contrastive learning loss, defined in Eq. (14), from PLHN, the segmentation DSC marginally decreased to 80.4$\%$. This outcome underscores the pivotal role that prototype-guided segmentation prediction plays within our proposed framework, leading to an approximate 1$\%$ enhancement in DSC. The empirical findings validate the proposition that the prototype-guided prediction strategy is more efficacious than solely relying on contrastive learning guided methodologies.\par
\begin{table}[t]\small 
\newcommand{\tabincell}[2]{\begin{tabular}{@{}#1@{}}#2\end{tabular}}
\centering
\renewcommand\arraystretch{1}
\setlength{\tabcolsep}{0.7mm}
\caption{The segmentation performance achieved by various methods for contrastive learning is evaluated in terms of DSC($\%$), PPV($\%$), SEN($\%$) and ASD(mm) with 95$\%$ confidence intervals on the internal breast tumor segmentation test dataset.}
\begin{tabular}{c|cccc}
\toprule[1.2pt]
Methods & DSC (\%) $\uparrow$  & PPV (\%) $\uparrow$  &SEN (\%) $\uparrow$  & ASD (mm) $\downarrow$ \\
\hline
HN+IDEAL \cite{basak2023ideal} &79.1$\pm$2.6 &\textbf{85.0$\pm$2.6} &78.1$\pm$2.7 &7.4$\pm$3.2 \\
HN+UPCoL \cite{lu2023upcol} &80.1$\pm$2.4 &79.8$\pm$2.5 &\textbf{85.2$\pm$2.3} &8.9$\pm$2.0 \\
\hline
PLHN wo CL &80.4$\pm$2.4 &84.7$\pm$2.2 & 80.7$\pm$2.6 &7.5$\pm$2.7\\
\textbf{PLHN (Ours)} &\textbf{80.6$\pm$2.3} &82.7$\pm$2.2 & 83.3$\pm$2.3 &\textbf{7.3$\pm$2.9}\\
\hline
\toprule[1pt]
\end{tabular}
\label{t:con}
\vspace{-0.3cm}
\end{table}
\vspace{-0.3cm}
\subsection{Hyper-parameter Analysis for Hybrid Network}
\indent Firstly, the parameters, including the feature dimension, $M$, the hidden size, $H_s$, and the number of transformer layers, $T$, are evaluated on the validation dataset. In this experiment, the prototype guided prediction module is not activated, and all the networks are optimized within 300 epoches within one stage. It is found in Table \ref{t:ab1} that larger values of $M$, $T$, or $H_s$ do not necessarily generate better segmentation performance. For example, the segmentation performance may not increase correspondingly when $M$ increases from 16 to 64. This is primarily because the required optimization epochs increase as the model complexity becomes heavier. Therefore, the segmentation performance may not necessarily increase within 300 epochs. Finally, the parameters $M$=32, $T$=8 and $H_s$=256 achieves the best performance of 79.1$\%$ on the validation dataset. These parameters are set as the default settings in PLHN.\par
\begin{table}[t]\small 
\small
\renewcommand{\arraystretch}{1}
\centering
\setlength{\tabcolsep}{0.24mm}
\caption{Segmentation performance achieved by different network parameters on the validation set for breast tumor segmentation is reported. It is noted that the prototypes guided module is not used in the experiment. The patches with input size $128 \times 128 \times 128$ are fed into the network for testing.}
\begin{tabular}{l|l|l|l|l|l|l|l|l|l}
\toprule[1.2pt]
Setting &1&2&3&4&5&6&7&8&9\\
\hline
$M$ &64&64&64&32&32&32&16&16&16\\
$H_s$ &384&256&192&384&256&192&384&256&192\\
$T$ &12&8&4&12&8&4&12&8&4\\
Flops (G) &700.3&638.7 &620.0 &265.1 &203.5 &184.8 &150.7 &89.2 &70.6 \\
Parameters (M) &30.2&15.2&10.6 &25.0 &10.0 &5.4 &23.7 &8.7 &4.1 \\
\hline
DSC &78.5&78.2& 78.3& 78.6& \textbf{79.1}& 78.2& 78.3& 77.9& 77.7\\
\toprule[1pt]
\end{tabular}
\label{t:ab1}
\vspace{-0.2cm}
\end{table}
\begin{table}[t] 
\newcommand{\tabincell}[2]{\begin{tabular}{@{}#1@{}}#2\end{tabular}}
\centering
\renewcommand\arraystretch{1.1}
\setlength{\tabcolsep}{2.2mm}
\caption{Segmentation performance achieved by different weights in optimization on the validation dataset for breast tumor segmentation.}
\begin{tabular}{l|l|l|l|l|l}
\toprule[1.2pt]
Settings  & {\bf $\lambda_1$=0.2}& {\bf $\lambda_1$=0.4} & {\bf $\lambda_1$=0.6} & {\bf $\lambda_1$=0.8}& {\bf $\lambda_1$=1}\\
\hline
$\lambda_2$=0  &80.5 &80.4 &80.4 &80.2  & 79.7 \\
$\lambda_2$=0.05 &\textbf{80.8} &~~$\backslash$ &~~$\backslash$ &~~$\backslash$  &~~$\backslash$ \\
$\lambda_2$=0.1  &80.6 &~~$\backslash$ &~~$\backslash$ &~~$\backslash$  &~~$\backslash$ \\
$\lambda_2$=0.15  &80.6 &~~$\backslash$ &~~$\backslash$ &~~$\backslash$  &~~$\backslash$ \\
\toprule[1pt]
\end{tabular}
\label{t:wei}
\vspace{-0.4cm}
\end{table}
\indent The cross-validation experiment for selecting weight for fusion output $\lambda_1$ and weight for contrastive loss $\lambda_2$ is displayed in Table \ref{t:wei}. It is observed that the parameters $\lambda_1$=0.2 and $\lambda_2$=0.05 achieve the best segmentation performance as 80.8$\%$ on the validation dataset. It has also been noted that increasing weight $\lambda_2$ for contrastive loss not necessarily increases the segmentation performance. Corresponding to the results in Table \ref{t:con}, the contrastive loss contributes to around 0.3$\%$ DSC improvement in PLHN, while the proposed prototype guided prediction strategy plays an more important role in boosting segmentation accuracy.\par
\vspace{-0.3cm}
\subsection{Effectiveness of Prototype Learning Guided Prediction}
\indent Firstly, the evaluation of two parameters, namely the number of categories ($C$) and the number of prototypes per category ($K$), is conducted on the validation dataset. For $C$=2, two categories, namely breast tumor and background, are defined. It is important to note that when $K$=1, each category corresponds to a single prototype, which is directly estimated through online clustering. This baseline achieves a DSC of 79.4$\%$ as shown in Table \ref{t:tproto1}. By increasing the number of prototypes from 1 to 3 and then to 5 ($K$:1$\rightarrow$3$\rightarrow$5), the segmentation performance gradually improves from 79.4$\%$  to 80.5$\%$  and finally to 80.8$\%$. These results confirm the effectiveness of prototypes. However, as the prototype number increases (i.e., $K$:5$\rightarrow$10$\rightarrow$15), the segmentation performance tends to saturate or even produce negative gains due to over-parameterization. When $C$=3, three categories are defined as breast tumor, breast, and background. It is observed that a DSC of 80.3$\%$ is obtained when three prototypes are used for each category. However, the performance exhibits slight changes with increasing $K$. Finally, the best performance on the validation dataset is achieved when $C$=2 and $K$=5. Therefore, these values are empirically chosen as the default parameters for the proposed approach.\par
\begin{figure}[t]
\centering
\centerline{\includegraphics[width=8.5cm]{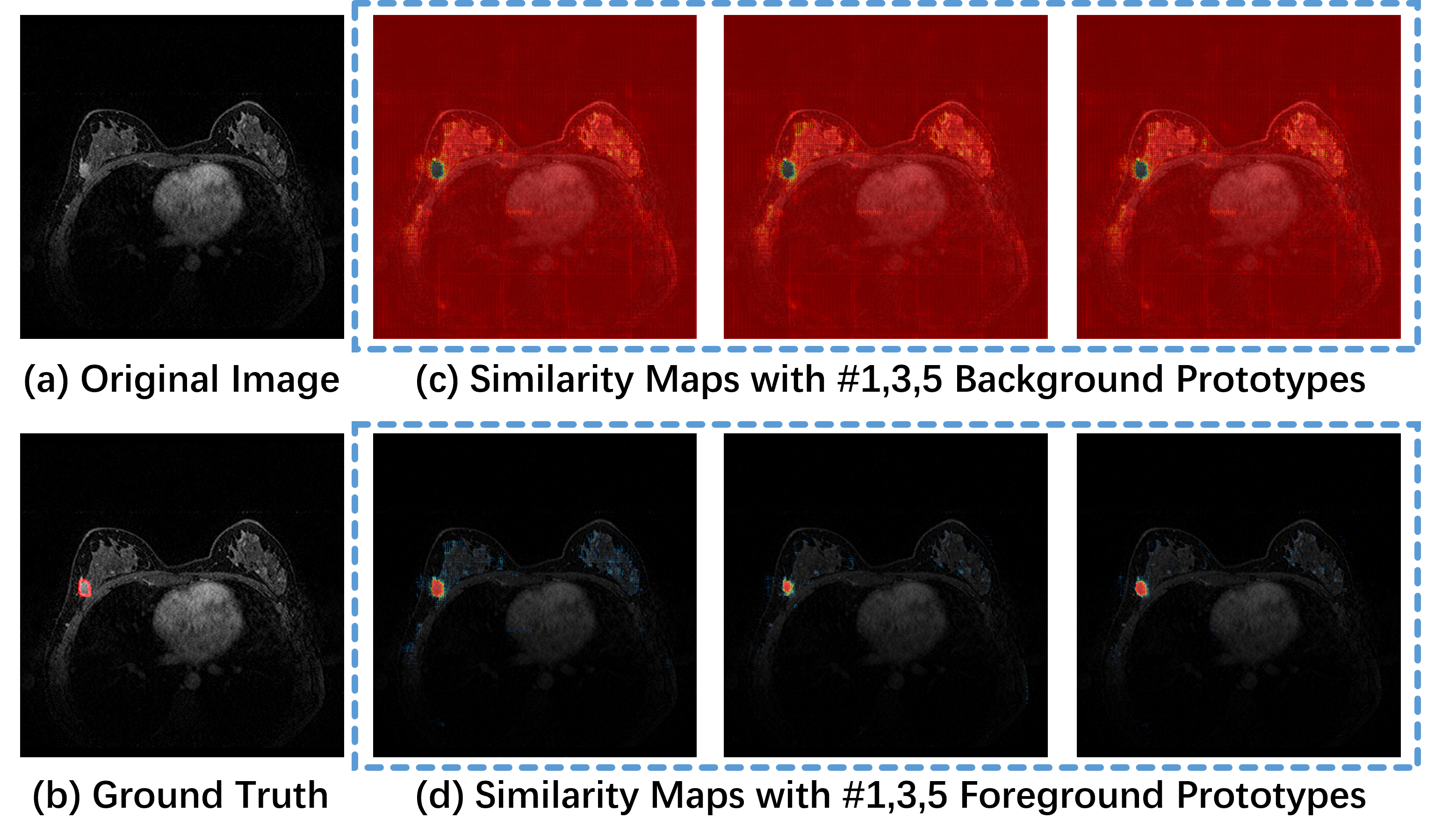}}
\caption{Illustration of the similarity maps highlighted by different prototypes. Each heatmap corresponds to the similarity values between a particular prototype with image feature $X$. Brighter pixels indicate higher similarity values. (a) is the original image. (b) is the ground truth segmentation mask. (c) are the similarity maps highlighted by background prototypes. (d) are the similarity maps highlighted by foreground prototypes.}
\label{fig:proto}
\vspace{-0.2cm}
\end{figure}
\begin{table}[t] 
\newcommand{\tabincell}[2]{\begin{tabular}{@{}#1@{}}#2\end{tabular}}
\centering
\renewcommand\arraystretch{1.1}
\setlength{\tabcolsep}{1mm}
\caption{Segmentation performance achieved by different number of prototypes and number of classes on the validation dataset for breast tumor segmentation.}
\begin{tabular}{l|l|l|l|l|l|l}
\toprule[1.2pt]
Settings &{Category} & {\bf K=1}& {\bf K=3} & {\bf K=5}& {\bf K=10}& {\bf K=15}\\
\hline
C=2 &tumor,other &79.4 &80.5 &\textbf{80.8} &80.4  & 80.6 \\
C=3 &tumor,breast,other &79.7 &80.3 &80.5 & 80.5 & 80.3 \\
\toprule[1pt]
\end{tabular}
\label{t:tproto1}
\vspace{-0.2cm}
\end{table}
\indent Secondly, the similarity maps between the decoder's normalized output features $X$ and the prototypical features $\mu$ are depicted in Fig.~\ref{fig:proto}. It can be observed that when $C=2$, the $K$ prototypes associated with the foreground are capable of effectively distinguishing tumor voxels from other tissues. Additionally, it has been discovered that different prototypes possess the ability to accentuate specific tissue regions in MRI images. For instance, the background prototypes tend to highlight non-tumor tissues, while the various foreground prototypes can emphasize a wide range of distinct and complementary shapes of breast tumors. This indicates that the acquired prototypes can function as reliable guidance for breast tumors segmentation.\par
\begin{figure}[t]
\centering
\centerline{\includegraphics[width=8.5cm]{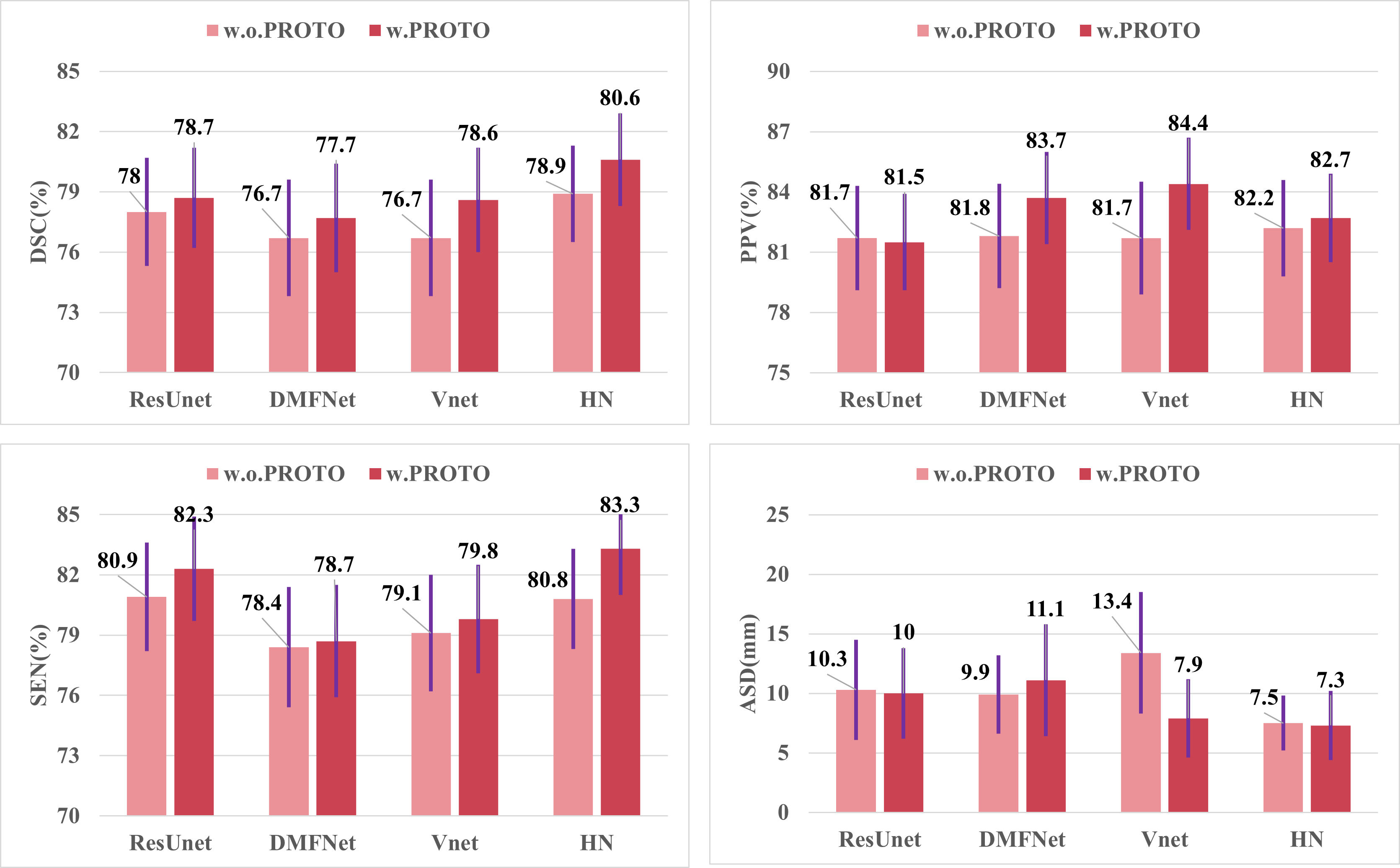}}
\caption{The segmentation performance of breast tumor achieved by various methods using a prototype learning-based prediction strategy is evaluated in terms of DSC($\%$), PPV($\%$), SEN($\%$) and ASD(mm) on the internal breast tumor segmentation test dataset.}
\label{fig:table7}
\vspace{-0.1cm}
\end{figure}

\indent Finally, to assess the generalizability of the proposed prototype guided prediction module, three baseline networks, namely Vnet, DMFNet, and ResUnet, were optimized jointly with prototypes. The extensive experimental results in Fig. \ref{fig:table7} have substantiated the efficacy of the prototype learning guided prediction strategy. For instance, if the prototype guided prediction module is integrated into the segmentation network, a 1.9$\%$ increase in DSC (78.6$\%$ vs 76.7$\%$) is achieved for Vnet and 1.0$\%$ DSC improvement (77.7$\%$ vs 76.7$\%$) is achieved by DMFNet. Moreover, around 0.7$\%$ increase in DSC (78.7$\%$ vs 78.0$\%$) is achieved by ResUnet. For the proposed hybrid network, nearly 1.7$\%$ increase in DSC (80.6$\%$ vs 78.9$\%$) is produced.\par
\vspace{-0.3cm}
\subsection{Ablation Study}
\begin{table} 
\newcommand{\tabincell}[2]{\begin{tabular}{@{}#1@{}}#2\end{tabular}}
\centering
\renewcommand\arraystretch{1}
\setlength{\tabcolsep}{1.1mm}
\caption{Segmentation performance of ablation study for the hybrid network. Epoch: the optimization epoches. E1+D: Encoder subnetwork-1 and Decoder. Trans: Transformer layers. E2+Trans: Encoder subnetwork-2 with transformer layers. TS: Two-stage optimization strategy. Proto: Prototypes guided prediction. Fus: Fusion for prediction.}
\begin{tabular}{l|l|l|l|l|l|l|l}
\toprule[1.2pt]
Epoch &E1$+$D &Trans &E2+Trans &TS  &Proto &Fus &DSC \\
\hline
300&~$\surd$& & & & & &77.5$\%$ \\
300&~$\surd$&$\surd$&& & & &77.1$\%$ \\
300&~$\surd$& &$\surd$ &  & & &78.9$\%$ \\
500&~$\surd$&$\surd$&& & & &78.0$\%$ \\
500&~$\surd$& &$\surd$ &  & & &\textbf{79.1$\%$} \\
\hline
500&~$\surd$&$\surd$&  &$\surd$ &  & &78.6$\%$ \\
500&~$\surd$& &$\surd$ &$\surd$ & &  &\textbf{79.4$\%$} \\
\hline
500&~$\surd$&$\surd$& &$\surd$ & $\surd$ & &79.5$\%$ \\
500&~$\surd$& & $\surd$&$\surd$ &$\surd$&  &80.2$\%$ \\
500&~$\surd$& &$\surd$ &$\surd$ &$\surd$  &$\surd$ &\textbf{80.6$\%$} \\
\toprule[1pt]
\end{tabular}
\label{t:t20}
\vspace{-0.3cm}
\end{table}
The segmentation performance of ablation study for the hybrid network with Encoder subnetwork-1 is evaluated and presented in Table \ref{t:t20}. It is noted that in the setting of Proto, the fusion strategy described in Eqs. (8) and (9) is not activated, the concatenated feature $cat[\Pi,S]$ will be applied for segmentation as Eq. (9) did directly.
\subsubsection{The Architecture of Two Encoder}
It is worth noting that when only the basic encoder-decoder architecture \lq\lq E1+D\rq\rq~is used, the achieved DSC is 77.5$\%$. However, when the transformer layers (Trans) are integrated into the hybrid network in the first stage of 300 epochs, the DSC of architecture \lq\lq E1+D+Trans\rq\rq ~drops to 77.1$\%$. This indicates that more optimization epochs are required when the transformer layers are incorporated. For example, the segmentation performance of architecture \lq\lq E1+D+Trans\rq\rq~is improved to 78.0$\%$ within 500 epoches. Moving forward, the segmentation performance of the hybrid network with parallel encoder subnetworks is evaluated. Once the encoder subnetwork-2 with transformer layers is utilized in architecture \lq\lq E1+D+E2+Trans\rq\rq, the DSC improves to 78.9$\%$ within 300 epochs during the first stage of optimization, which is far better than 77.1$\%$ of the single branch network. This suggests that parallel encoder subnetworks can enhance the optimization efficiency. Once optimized for another 200 epoches, the segmentation DSC of architecture \lq\lq E1+D+E2+Trans\rq\rq~can be improved to 79.1$\%$. This can be primarily attributed to enhanced capacity achieved through the parallel encoders' ability to extract diverse features.
\subsubsection{Two Stage Optimization}
By implementing the two-stage optimization strategy (TS), performance of the \lq\lq E1+D+Trans\rq\rq~architecture can be enhanced from 78.0$\%$ to 78.6$\%$ after 500 epoches. Furthermore, an additional 0.3$\%$ improvement in DSC is achieved for the \lq\lq E1+D+E2+Trans\rq\rq~ architecture with the implementation of the two-stage optimization (79.4$\%$ vs 79.1$\%$). The experimental findings suggest that the two-stage optimization strategy, which involves pretraining the parameters of the backbone network initially, serves as an effective method for enhancing PLHN's optimization.
\vspace{0cm}
\subsubsection{Effectiveness of Prototypes}
Moreover, when the Prototype Guided Prediction (Proto) module is optimized jointly with the backbone network, the architecture \lq\lq E1+D+Trans+Proto\rq\rq~generates a DSC as 79.5$\%$, thus demonstrating 0.9$\%$ DSC gain. If the Proto module is combined with the two-stage optimization strategy, the DSC of architecture \lq\lq E1+E2+Trans+Proto\rq\rq~is elevated to 80.2$\%$ within 500 epochs, 0.7$\%$ better than that of single encoder architecture. Ultimately, the incorporation of the attention-based fusion module (Fus) into the hybrid network leads to the highest DSC as 80.6$\%$. In fact, if the prototype based non-parametric prediction strategy in \cite{zhou2022rethinking} is applied for breast tumor segmentation directly, the DSC is only 72.5$\%$, indicating that fusing prototype similarity with backbone features is more effective for breast tumor segmentation.\par
In conclusion, the experimental findings provide substantial evidence that proposed hybrid network, which combines parallel encoders and prototype guided prediction strategy, achieves exceptional performance in breast tumor segmentation.\par
\begin{table}[t]\small 
\newcommand{\tabincell}[2]{\begin{tabular}{@{}#1@{}}#2\end{tabular}}
\centering
\renewcommand\arraystretch{1}
\setlength{\tabcolsep}{0.7mm}
\caption{Computation cost of different methods measured by Parameters(M) and FLOPs(G). The patches with input size $128 \times 128 \times 48$ are fed into the network for testing.}
\begin{tabular}{l|l|l|l|l}
\toprule[1.2pt]
Methods &Vnet\cite{milletari2016v}& ResUnet\cite{yu2017volumetric} &DMFNet\cite{chen20193d} &MHL\cite{zhang2018hierarchical}\\
\hline
Parameters (M) &9.5 & 11.2 &3.9 &1.7 \\
FLOPs (G) &37.3 & 152.2 &9.9 &357.0 \\
DSC &76.7$\%$ &78.0$\%$ &76.7$\%$ &77.1$\%$ \\
\hline
Methods &MTLN\cite{zhang20213d} &UXNET \cite{lee20223d} &ALMN\cite{zhou2022three} &\textbf{PLHN} \\
\hline
Parameters (M) &9.2 &4.1 &53.6 &11.0  \\
FLOPs (G) &92.0 &64.5 &2478.8 &146.1  \\
DSC &77.6$\%$ &77.3$\%$ &79.3$\%$ &\textbf{80.6$\%$} \\
\toprule[1pt]
\end{tabular}
\vspace{-0.3cm}
\label{t:t22}
\end{table}
\vspace{-0.3cm}
\subsection{Analysis of Network Complexity}\label{exp:c}
The complexity of various network architectures for breast tumor segmentation is assessed and compared, at the input size as $128 \times 128 \times 48$. The computation costs and segmentation DSC of different methods are listed in Table \ref{t:t22}. The primary observation is that more sophisticated networks yield better segmentation performance. For instance, the second best method, ALMN, achieves a segmentation DSC of 79.3$\%$,
but its computation cost is significantly higher compared to other methods under consideration. Specifically, ALMN has 53.6$M$ parameters and 2478.8$G$ FLOPs (Floating-Point Operations), which is nearly 250 times larger than those of the compared method DMFNet (3.9$M$ parameters and 9.9$G$ FLOPs). Nevertheless, ALMN demonstrates a noteworthy 2.6$\%$ DSC improvement in comparison to DMFNet. On the contrary, the proposed PLHN achieves a better balance between computation costs and segmentation accuracy. For instance, the FLOPs of PLHN amount to 146.1$G$ and the parameters tally up to 11.0$M$, rendering it lighter than ResUnet, MHL, and ALMN. Despite being lighter, PLHN manages to attain superior segmentation performance. Conversely, other compared methods including Vnet, DMFNet, MTLN, and UXNET exhibit lower computation costs. However, when these lighter network architectures are employed, the segmentation performance has the potential to plummet precipitously. In summary, PLHN accomplishes a minimum of 3$\%$ increase in DSC, relative to the aforementioned methods. It is noted that the FLOP of hybrid network without prototype learning based module is only 76.3$G$ but it still achieves better segmentation performance as 79.4$\%$ than compared methods. Notably, even if DMFNet and Vnet may also be optimized through the utilization of the prototype learning guided optimization strategy (as shown in Fig \ref{fig:table7}), PLHN still achieves a 2.9$\%$ and 2$\%$ increase in DSC when compared with these two methods, respectively.
\begin{table}\small 
\newcommand{\tabincell}[2]{\begin{tabular}{@{}#1@{}}#2\end{tabular}}
\centering
\renewcommand\arraystretch{1}
\setlength{\tabcolsep}{1mm}
\caption{The diagnosis performance using different segmentation masks. The results corresponds to metrics such as AUC, Accuracy, Precision, Recall and F1-score are reported. The best two results for each criteria are indicated in blod.}
\begin{tabular}{l|l|l|l|l|l|l}
\toprule[1.2pt]
Methods & GT &ResUnet & MHL &Tumorsen &ALMN &PLHN \\
\hline
AUC (\%) $\uparrow$ &\textbf{68.8} &63.2 &65.7 &65.7 &\textbf{70.9} &66.6\\
Accuracy (\%) $\uparrow$ &\textbf{64.1} &61.1 &61.0  & 61.3 &63.6 &\textbf{63.8} \\
Precision (\%) $\uparrow$ &\textbf{66.4} &62.0 &58.9 &57.6 &\textbf{70.6} &64.0\\
Recall (\%) $\uparrow$ &\textbf{33.2} &26.8 &24.9 &32.6 &24.4  &\textbf{33.5} \\
F1-score (\%) $\uparrow$&\textbf{42.5} &35.8 &32.8 &39.9 &34.7 &\textbf{41.3}\\
\hline
DSC (\%) $\uparrow$&* &84.4 &83.3  &84.0 &84.9 &\textbf{85.6}\\
\toprule[1pt]
\end{tabular}
\label{t:trad}
\vspace{-0.5cm}
\end{table}
\vspace{-0.3cm}
\subsection{Performance of HER2 Status Classification}
The classification performance between HER2-positive and HER2-negative cases was assessed by calculating various criteria, including the area under the receiver operating characteristic curve (AUC), accuracy, precision, recall, and F1-score, utilizing different methods. For this particular experiment, a total of 500 cases with HER2 labels were selected from MRI images obtained from GD-hospital. Among these cases, 214 were HER2-positive and 286 were HER2-negative. Additionally, other 565 cases were chosen from the dataset as the training images to optimize segmentation networks, specifically ResUnet, MHL, ALMN, Tumorsen, and PLHN. Subsequently, the MRI images were segmented using various methods. In the diagnosis pipeline, radiomics features were extracted from regions of interest (ROIs) using an in-house feature analysis program implemented in Pyradiomics (http://pyradiomics.readthedocs.io) \cite{van2017computational}. Then the logistic regression algorithm (LR) \cite{demaris1995tutorial} was employed for HER2 positive or negative classification. Given the limited number of training samples, 5-fold validation was conducted, and the average values of AUC, accuracy, precision, recall, and F1-score on the validation dataset were reported.\par
In addition to the ground truth tumor masks, the breast tumor masks generated by ResUnet, MHL, Tumorsen, ALMN and PLHN are additionally utilized for feature extraction, the DSC values for these methods on the selected 500 cases are as follows: 84.4$\%$, 83.3$\%$, 84.0$\%$, 84.9$\%$ and 85.6$\%$ for these five methods, respectively. It can be observed that when the ground truth tumor masks are utilized, all of the corresponding classification criteria rank within the top-2 positions in Table \ref{t:trad}. Notably, once the features extracted via tumor masks of PLMN are applied for
classification, the accuracy (63.8$\%$), recall (33.5$\%$) and F1-score (41.3$\%$) also ranked within top-2 positions in comparison.
Furthermore, our previous approach, ALMN, surpasses the other methods with the best Area Under the Curve (AUC) value of 70.9$\%$ and precision   (70.6$\%$). The classification results indicate the significant value of the segmentation masks generated by PLHN for better diagnosis performance. Compared to the methods utilizing features extracted from ground-truth tumor regions, PLHN achieves promising results in HER2 status classification, suggesting that the breast tumor regions generated by our automatic method are valuable for radiomic analysis.\par
\begin{figure}
\centering
\centerline{\includegraphics[width=9cm]{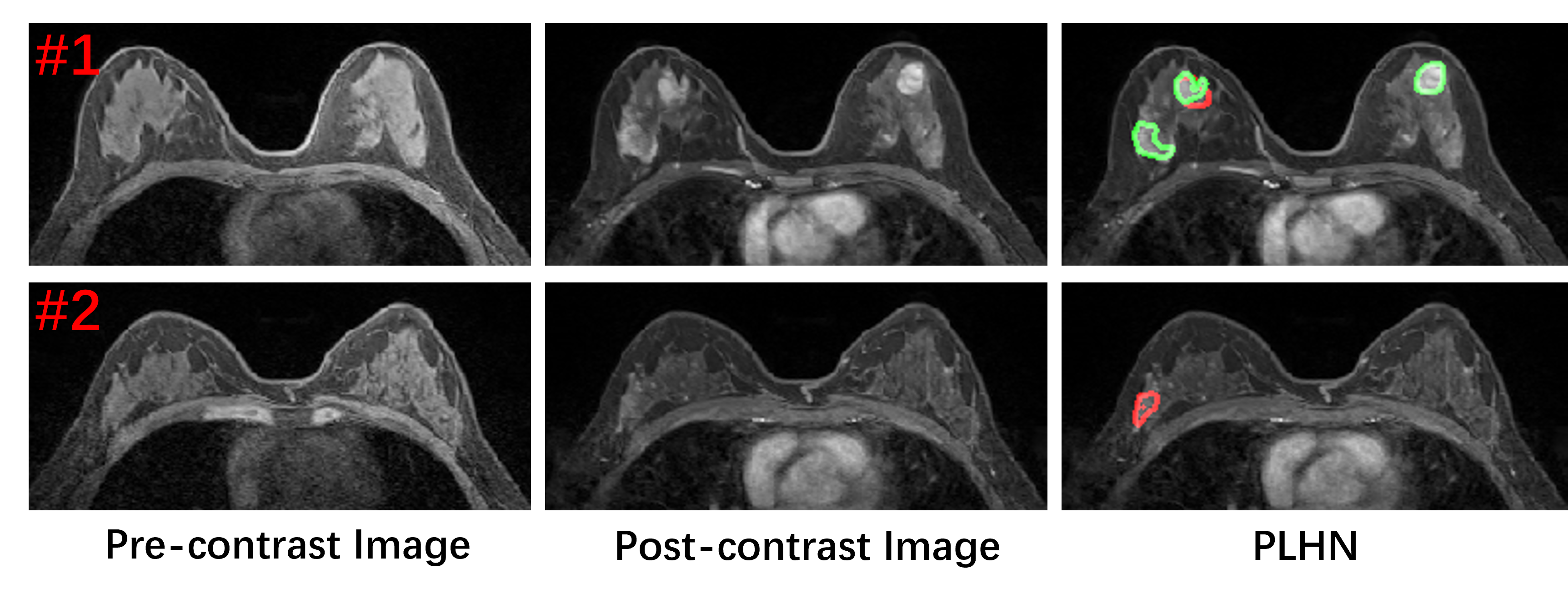}}
\caption{Illustration of two failure cases. The red and
green curves indicate manual annotations and network predictions,
respectively.}
\label{fig:fail}
\vspace{-0.3cm}
\end{figure}
\vspace{-0.3cm}
\section{Conclusion and Discussion}
\indent In this paper, we propose a prototype learning guided hybrid network (PLHN) to perform fine-grained 3D segmentation for breast tumors in MRI images. We first design a simple and effective hybrid network by combining convolution layers and transformer, in which the 3D transformer layers capture the global dependency between bottleneck features. To enhance the optimization efficiency of the hybrid network, we designed two parallel encoder subnetworks. Specifically, one encoder subnetwork extracts features for the skip connection to the decoder, and the other encoder subnetwork is designed to extract features for the transformer layers. To enhance the discriminating ability of hybrid network, a prototypes guided prediction module is designed by calculating the prototypical features for each category. Lastly, we propose a novel attention-based fusion module to merge the decoder's output features and the similarity maps with prototypical features, thus generating breast tumor masks with increased accuracy. Experimental results derived from internal and external DCE-MRI breast tumor segmentation datasets demonstrate the superior performance of the proposed PLHN when compared to other state-of-the-art (SOTA) segmentation methods.\par
The limitation of PLHN for segmentation is depicted in Fig. \ref{fig:fail} via two typical instances of failure. In the first scenario, certain tissues resembling breast tumors are erroneously classified as abnormal. This can be attributed to the fact that the subtraction image between the pre-contrast and post-contrast images may not adequately highlight numerous normal voxels. Conversely, in the second case, the breast tumors fail to be accurately segmented due to the absence of recognizable cues when examining the difference map between these two types of images. Once the DCE-MRI images are utilized for breast tumor diagnosis, the delineation of tumor shape can be achieved by comparing signal variation across different image phases. If the variation between the pre-contrast and post-contrast images is negligible, the tumor voxels may not be effectively distinguished. In future research, we intend to leverage multiple post-contrast phases, as suggested by \cite{zhang2023robust}, to further enhance the segmentation performance. Moreover, comparing the pre-contrast and post-contrast images plays an important role in PLHN. However, the performance of PLHN may degrade when only the pre-contrast images are available. In the future, we also aim to address this limitation by exploiting diffusion-based synthesis techniques to enable flexible inputs.\par
Based on the segmentation masks, we have demonstrated that the automatically generated tumor masks can be employed in radiomics to identify the HER2-positive subtype from a HER2-negative status with comparable accuracy to the analysis results obtained from manual tumor segmentation. It is noted that solely using radiomics feature is not the gold standard to classify HER2 status. For example, a novel convolutional neural network approach is proposed \cite{2022Deep} on whole slide images to predict HER2 status with increased accuracy. In our future work, we will explore utilizing CNNs to improve classification accuracy of HER2 status.\par
In summary, the suggested segmentation framework PLHN significantly improves segmentation efficiency and establishes practicality for the subsequent diagnosis task. We firmly believe that this research represents a crucial step towards enhancing the clinical diagnosis of breast cancer.
\bibliographystyle{ieeetr}
\bibliography{refs}
\end{document}